\documentclass[12pt]{article}
\usepackage{amsfonts}
\usepackage{amsmath}
\usepackage{amsthm} 
\usepackage{graphicx}
\usepackage{float}
\usepackage{enumerate}

\usepackage{hyperref}

\newcommand{\NoBlackBoxes}{\global\overfullrule0pt}
\NoBlackBoxes

\newcommand{\E}{{\mathbb E}}
\newcommand{\Var}{{\mathbb Var}}

\newcommand{\Pa}{{\mathbb P}}

\newcommand{\C}{{\mathbb C}}
\newcommand{\R}{{\mathbb R}}
\newcommand{\N}{{\mathbb N}}

\newcommand{\Ncal}{{\mathcal N}}

\newcommand{\Bphi}{{\boldsymbol\phi}}
\newcommand{\Blambda}{{\boldsymbol \lambda}}
\newcommand{\BSigma}{{\boldsymbol\Sigma}}
\newcommand{\BGamma}{{\boldsymbol\Gamma}}

\newcommand{\Bl}{{\boldsymbol \ell}}

\newcommand{\Bh}{{\mathbf h}}

\newcommand{\BL}{{\mathbf L}}
\newcommand{\Bx}{{\mathbf x}}
\newcommand{\Bu}{{\mathbf u}}
\newcommand{\Bv}{{\mathbf v}}
\newcommand{\BX}{{\mathbf X}}
\newcommand{\BD}{{\mathbf D}}
\newcommand{\BK}{{\mathbf K}}
\newcommand{\BZ}{{\mathbf Z}}
\newcommand{\BY}{{\mathbf Y}}
\newcommand{\Bone}{{\mathbf 1}}
\newcommand{\BoneT}{{\mathbf 1^{\boldmath\perp}}}
\newcommand{\Bzero}{{\mathbf 0}}
\newcommand{\BSigmaL}{{\mbox{\boldmath$\Sigma^L$}}}

\renewcommand{\proof}{\noindent {\em Proof: }}
\renewcommand{\qed}{{\hfill $\square$}}

\newcommand{\tr}{{\rm tr }}

\renewcommand{\omega}{\varpi}

\newcommand{\Ksa}{{G}}

\newcommand{\qL}{q}

\newcommand{\LL}{L}
\newcommand{\XX}{X}

\newcommand{\scp}[2]{\langle #1,#2\rangle}

\newcommand{\II}{K}

\newcommand*\rfrac[2]{{}^{#1}\!/_{#2}}

\newcommand{\one}{{1\!\!1}}
\newcommand{\VaR}{{\rm VaR}}

\newcommand{\ES}{{\rm ES}}

\newcommand{\myfootnote}{\footnote}

\newtheorem{proposition}{Proposition}
\newtheorem{lemma}[proposition]{Lemma}
\newtheorem{theorem}[proposition]{Theorem}

\newtheorem{corollary}[proposition]{Corollary}
\newtheorem{remark}[proposition]{Remark}
\newtheorem{exampleemph}[proposition]{Example}   
\newtheorem{foo}[proposition]{Remarks}

\begin{document}

\title{Economic Neutral Position: How to best replicate not fully replicable liabilities?
}
\author{Andreas Kunz,  Markus Popp \myfootnote{Munich Re. Letters: K\"oniginstrasse
107, 80802 M\"unchen, Germany. Emails: \{akunz,mpopp\}@munichre.com.}
}

\date{Version: \today}

\maketitle

\abstract{
Financial undertakings often have to deal with liabilities of the form ``non-hedgeable claim size times value of a tradeable asset", e.g.~foreign property insurance claims times fx rates.
Which strategy to invest in the tradeable asset is risk minimal? 
We generalize the Gram-Charlier series for the sum of two dependent random variable, which allows us to
expand the capital requirements based on value-at-risk and expected shortfall. 
We derive a stable and fairly model independent approximation of the risk minimal asset allocation  in terms of the claim size distribution and the moments of asset return.
The results enable a correct and easy-to-implement modularization of capital requirements into a market risk and a non-hedgeable risk component.\\

\noindent
{\bf Keywords:} risk measure; risk minimal asset allocation; incomplete markets; modular capital requirements; perturbation theory;
Gram-Chalier series; Cornish-Fisher quantile approximation; quantos;
Solvency II; standard formula; SCR; market risk; internal model; replicating portfolio; \\

\noindent
{\bf JEL Classification:} D81; G11; G22; G28;
}

\section{Introduction}
\label{sec:Intro}

We consider a liability of product structure $\sum_i \LL_i \cdot \XX_i$, where $\XX_i$ are hedgeable risk factors and $\LL_i$ represent
stochastic notionals or claim sizes that are not replicable by financial instruments.
It is well known that such liability is not perfectly replicable, since the number of risk drivers exceeds the number of involved
hedgeable capital market factors.

This liability structure is of high practical relevance.
Prominent examples stem from insurance:  $\LL_i$ denoting the claims from property insurance portfolios in foreign currencies and $\XX_i$ denoting the exchange rates, or, $\LL_i$ the benefit payments of pure endowment policies staggered by maturities (depending on realized mortality) and $\XX_i$ the risk-free discount factors.
Also for the banking industry such liability structure is relevant, in particular for measuring 
 the credit value adjustment (CVA) risk for non-collateralized derivatives with counterparties for which  no liquid credit default swaps exists: e.g.~the CVA for a non-collateralized commodity forward contract
can be written in the above structure with $\LL_i$ denoting the default rate of the counterparty in the time interval $t_i$ (multiplied by the loss-given-default ratio) and
$\XX_i$ denoting a commodity call option expiring at $t_i$.
The latter represents the loss potential due to counterparty default at $t_i$
in case of increasing commodity prices.\footnote{Hereby we assumed independence of the default rates from the credit exposure against the counterparty due to an increase of the commodity forward rates beyond the pre-agreed strike, refer e.g. to \cite{Gregory} for details.}

To which extent can the risk from the above liability structure be mitigated by
trading in the capital market factors $\XX_i$?
The residual risk must be warehoused and backed with capital. The capital requirement for a financial institution is obtained in theory by applying a risk measure $\rho$ on the distribution of its surplus (i.e.~excess of the value of assets over liabilities) in one year, which is the typical time horizon for risk measurement.
Hence we aim to find the optimal strategy to invest in the assets $\XX_i$ that minimizes the capital requirements.
Intuition tells us that investing more than the expected claim size into the respective hedgeable asset $\XX_i$ makes sense, since
large liability losses are usually driven by events where both the claim sizes and the asset values develop adversely.
As risk measures focus on tail events, the excess investments in $X_i$ mitigate that part of the liability losses that stems from an increase in $\XX_i$.
The essential task now is to quantify this excess amount.

Without loosing too much of generality we assume that $\LL_i$ and $\XX_j$ are pairwise \textit{independent} for any combination of $i$ and $j$
and that there is \textit{no continuous increase in information} concerning the states of  $L_i$ during the risk measurement horizon.
The latter assumption is almost tantamount to the assumption that claim sizes $\LL_i$ are not hedgeable. As a consequence there is no need to adjust the holdings in $\XX_i$ dynamically within the year.
If $\LL_i$ and $\XX_j$ were not independent, then in most practical applications $\LL_i$ can be expressed by regression techniques
as a function 
of the capital market factors $\XX_j$ plus some residual $\LL_i'$ which then is independent of all $\XX_j$ by construction.

Even if the $\XX_i$ and $\LL_j$ are normally or log-normally distributed,
the derivation of the risk minimal asset allocation is not straight forward, since
products of log-normal variables are again log-normal but sums are not and vice versa for normal variables.

This paper is to be interpreted in the context of hedging in incomplete markets. The results relate to the approach of quantile hedging or efficient hedging initiated by F\"ollmer \& Leukert \cite{FoellmerLeukert_QuantileHedging} and
\cite{FoellmerLeukert_EfficHedg} and extended in particular by Cvitanic \& Spivac \cite{CvitanicSpivac99}, Cvitanic \cite{Cvitanic2000},  Cvitanic \& Karazas \cite{CvitanicKarazas2001} and Pham\cite{Pham}, see also chapter 8 of  F\"ollmer \& Schied \cite{FoellmerSchied} and the reference therein.
For a given budget constraint on the hedge, the (static) quantile hedging strategy results
 for a liability of product structure as described above
in holding a certain amount of the tradeable asset, which corresponds to the distribution of the non-hedgeable claim size distribution truncated at a particular quantile. The efficient hedging framework provides some determining conditions for that truncation level.
Similar conditions are derived also when the shortfall risk of failing to (over-)hedging the liability is minimized instead of the probability.
The results of this paper allow to approximate this truncation level explicitly in terms of characteristics of the claim size and asset distributions.

Another approach to hedging in incomplete markets is mean variance hedging or -- more specifically -- (local) quadratc risk-minimizing strategies initiated by F\"ollmer \& Sondermann \cite{FoellmerSondermann} and developed further by F\"ollmer \& Schweizer \cite{FoellmerSchweizer} and Schweizer \cite{Schweizer01aguided}.
Applications of these techniques to insurance mathematics have been intensively studied in particular by  M{\o}ller \cite{Moller_RiskMinUnitLinked}, \cite{Moller_RiskMinInsur}, \cite{Moller01_HedgingEqtLinked} and \cite{Moller01_TrafoActuarPricip}.
Here the insurance risk process (stochastic mortality) is time-continuous and hence reveals a dynamic hedging strategy that reacts immediately to insurance risk changes. As the variance of the hedging error is minimized instead of a down-side focussed risk measure, the replication is always based on the current expectation of the insurance risk factor (mortality), i.e.~no overhedging of the best-estimate claim size by a specific fraction of the pure insurance risk occurs as in our approach. Moreover the hedging risk is minimized under the risk-neutral measure and not under the physical measure that is relevant for risk measurement.
A further approach towards hedging of insurance claims in an incomplete market is the utility indifference pricing approach initiated by Schweizer \cite{Schweizer01_FromActuarToFinValuPrincip} and Becherer \cite{Becherer}, refer also to
M{\o}ller \cite{Moller03_IndiffPricInsur},  Henderson \& Hobson \cite{HendersonHobsonUtilIndiffPricOverview} and also
 to the survey paper Dahl \& M{\o}ller \cite{MollerDahl06_ValuHedgLifInsurSystemMortRisk} that combines utility indifference pricing with quadratic risk-minimization.

In this paper, we analyze 
the risk measures value-at-risk and expected shortfall.
Our first results concern 
 the {\em particular asset allocation}, i.e.~the initial holding in the asset $X$
which makes the capital requirements independent of the asset distribution.
We show  in section \ref{sec:GenResults} that
in the one-dimensional case
this particular asset allocation equals for both risk measures the value-at-risk of the non-hedgeable claim size distribution, i.e.~coincides with the capital requirement when the asset volatility tends to zero.
Moreover, this particular asset allocation is risk minimal in the expected shortfall case; the value-at-risk based  capital requirements on the other hand are still decreasing when less than this exceptional amount is invested in $\XX$.

In the second part of this paper we apply perturbation techniques to the capital requirements.
Classical expansion techniques such as the Gram-Charlier series (refer to \cite{Charlier} for the seminal paper)
approximate the distribution of a random variable in terms of its moments or cumulants.
Typically the Gaussian density is used as base function resulting in an expansion in terms of Hermite polynomials.
The Cornish-Fisher expansion (first published in \cite{CornishFisher}) uses a similar approach to expand the quantiles of random variables.
Similar to the Gram-Charlier series, the Edgeworth expansion \cite{Edgeworth} approximates the distance of the
sum of i.i.d.~random variables (properly scaled) to the Gaussian density, which is closely linked to the bootstrap method, refer to Hall
 \cite{Hall_EdgeworthExpansion}.
For details  on classical expansion techniques and further developments refer to the monographs Kolassa \cite{Kolassa_ApproxMethod_LecNotes}, Johnson et al. \cite{JohnsonKotz}, Wallace \cite{Wallace_AsymptApproxDistrib}, and the references therein.
These classical expansion techniques celebrate a revival in financial mathematics,  refer e.g.~to Ait-Sahalia et al. \cite{AitSahalZhanMykland_Edeworth} and the references therein.

A straight-forward application of the Cornish-Fisher approach to expand the value-at-risk of the surplus in terms of
Hermite polynomials fails to reproduce the distribution-independent relation 
at the particular asset allocation, which we derive in the first part of this paper. The reason is that due to the product structure of the liability
the distribution of the surplus becomes so irregular that the  quantile cannot be well approximated by the third and forth excess moments compared to the Gaussian distribution.
We prove in Proposition \ref{prop:ExpansSum2depRVs}
a Gram-Charlier-like expansion of the sum of two dependent random variables, where not the Gaussian density is used as base function but the distribution of one variable instead.

Writing the surplus as sum of a non-hedgeable term and a perturbation term based on the hedgeable assets, Proposition \ref{prop:ExpansSum2depRVs} yields an expansion of the surplus distribution in terms of moments of the hedgeable assets.
Expanding  in terms of the normal or log-normal asset volatility, we obtain an approximation of the
capital requirement (value-at-risk and expected shortfall based) up to forth order in the asset volatility (refer to
Theorem \ref{thm:PerturbVar} and Corollary \ref{cor:ES3rdOrdExpan1Dim}), which also results in an expansion of the optimal asset allocation.
The approach generalizes easily to the multivariate case where several assets and non-hedgeable claim sizes are involved; the second order expansion
of the capital requirements  in terms of asset volatility is presented in Theorem \ref{thm:PerturbVarMultiDim} (value-at-risk) and Corollary \ref{corr:PerturbESMultiDim} (expected shortfall). We show that the sum of the optimal investment amounts is given by the optimal  amount in
the associated univariate case; further, the allocation of the total optimal investment amount into the
single asset dimensions follows the covariance principle as long as the non-hedgeable claim sizes are multi-variate Gaussian (refer to Theorems \ref{corr:OptPhiAlloc} and
\ref{thm:AllocPhiOptLmultiNormal}).
Numerical studies show  that the derived expansions are stable even for large log-normal asset volatility levels.

Our results relate also to the replicating portfolio techniques, that have been recently studied with financial mathematical rigour, refer to the work of
Natolski \& Werner \cite{NatolskiWerner_RepPfMathAnalyDiffApproach},
Pelsser \& Schweizer \cite{PelsserSchweizer_RepPFDiffLSMC} and
Cambou \& Filipovi\'{c} \cite{CambouFilipovic_RepPF}.
The main focus of these papers is to analyze how to best approximate
complex not-perfectly hedgeable claims by investment strategies based on a specified investment universe (including derivatives);
this best approximating replicating portfolio is then used for measuring market risk.
Whereas the admissible financial claims are much more complex and general than  liabilities of product type (as  analyzed in our paper), the stochastic modelling of  insurance risk factors and the interaction of the insurance and financial stochastics is not explicitly analyzed.

To determine 
the asset allocation that minimizes capital requirements in a rather generic and model independent way is important for its own sake.
This objective is even more relevant for the modularization of
capital requirements into a capital market and a non-hedgeable risk component. This has become market standard since
deriving capital requirements via a joint stochastic modeling of all (hedgeable and non-hedgeable) risk factors turned out to be too complex.
The financial benchmark (Economic Neutral Position) against which the actual investment portfolio is measured to obtain the capital market risk component must obviously coincide with the risk minimal asset allocation. Our results show that the Economic Neutral Position replicates the financial risk factors of the liabilities on the basis of the expected claim size plus some safety margin.
Solvency II, the new capital regime for European insurers, does not recognize this safety margin
in the modularized Standard Formula approach, which can result in significant distortions of the total risk compared with the (correct) fully stochastic approach, refer to \cite{SolvII_Directive} for details.
The results of this paper provide a simple and stable approximation of the required safety margin in the Economic Neutral Position, such that the modularized capital requirement approach keeps its easy-to-implement property; e.g.~for non-hedgeable risks with normal tails the safety margin amounts to $85\%$ of the insurance risk component in the Solvency II context.

\section{Setup and Preliminary Results}
\label{sec:Setup}

Consider a financial undertaking whose {capital requirement is determined by applying a risk measure $\rho$ on
its surplus $S$ in one year.
The  value of the liabilities  at year one shall factorize in the form 
$\sum_{i=1}^n X_i \cdot L_i$, where the real-valued random variables $X_i$ and $L_i$ denote the value of a $i$-the tradeable asset and the claim size associated to this asset, respectively.
These variables live on a probability space with measure $\Pa$ together with a risk free numeraire investment (money market account).
The $X_i$ are assumed strictly positive and independent of  $L_j$, $i,j=1, \dots, n$. All financial quantities are expressed in units of the numeraire.

The financial undertaking can invest its assets with initial value $A_0\geq 0$  into the tradeable assets $X_i$ with initial value $x_i$ or into the numeraire.
We assume that  additional information concerning the claim sizes becomes known only at year one, i.e.~there is no continuous increase in information concerning the state of $L_i$ during the year.  Hence there is no need to adjust the holdings in $X_i$ dynamically within the year. We denote by $\phi_i \geq 0$ the number of units the financial undertaking invests statically into the asset $X_i$ as of today; the remaining asset value $A_0- \sum_{i=1}^n \phi_i\cdot x_i$ is invested into the numeraire.

We denote in the sequel column vectors and matrices in bold face,~e.g. $\Bphi$ is the column vector $(\phi_1, \dots, \phi_n)'$, where the prime superscript denotes the transposed vector or matrix, respectively. By $\scp{\cdot}{\cdot}$ 
we denote the scalar product.
The value of the surplus 
at year one 
is a function of the asset allocation $\Bphi$ 
and reads expressed in units of the numeraire
\begin{equation}\label{eq:DefSphiOrig}
S(\Bphi) := \sum_{i=1}^n\phi_i\cdot X_i + A_0 - \sum_{i=1}^n\phi_i\cdot x_i - \sum_{i=1}^n X_i\cdot L_i
= \scp{\BX - \Bx}{\Bphi} + A_0 - \scp{\BX}{\BL}\, .
\end{equation}
We analyze the risk measures value-at-risk $\VaR_{\alpha}$ and expected shortfall $\ES_{\alpha}$ at tolerance level $1\!-\!\alpha$ for some small $\alpha >0$. Typically $\alpha=0.01$ for banks and $=0.005$ for European insurance companies. Refer to \cite{FoellmerSchied} for details of the definition of $\VaR_{\alpha}$ and  $\ES_{\alpha}$.
We use the notation $\rho$ if the expression is valid for both analyzed risk measures $\rho \in \{\VaR_{\alpha},\ES_{\alpha}\}$.

We aim to find the optimal holdings $\Bphi^*$ in the tradeable assets that minimize the risk of the surplus, i.e.
$$\rho[S(\Bphi^*)] = {\min}_{\Bphi\in \R_+^n} \;  \rho[S(\Bphi)] \, .$$
Note that we do not allow for leverage, i.e.~$\phi_i < 0$ is forbidden.
We assume the following technical conditions:
\begin{eqnarray}
&&\mbox{$X_i$, $X_i^{-1}$, $L_i$, and $\scp{\BX}{\BL}$ are integrable for every $i=1,\dots, n$,}\label{eq:TechAssumpIntegr}\\ 
&&\mbox{$\BL$ has a bounded and strictly positive $n$-dimensional density $f_\BL$ .}
\label{eq:TechAssumpLdensity}
\end{eqnarray}
To simplify the minimization of $\rho[S({\Bphi})]$ we assume without loss of generality
\begin{equation}\label{Ass:WLOG_Assumpt}
      \E[\BX] = \Bx  = \Bone \, , \quad 
      \E[\BL] = {\mathbf 0} \, , \quad   A_0 = 0  \, ,
\end{equation}
where $\Bone$ and ${\mathbf 0}$ denote the column vector with all entries equal to one and zero, respectively.
The first assumption means in particular that $\BX$ is fairly priced. Further these assumptions imply that $S(\Bphi)$ has zero mean and hence reads
\begin{equation}\label{eq:DefSphiSimplif}
S(\Bphi)= \scp{\BX - \Bone}{\Bphi} - \scp{\BX}{\BL} \, .
\end{equation}
These simplifying assumptions can be justified
by centering and normalizing $S(\Bphi)$, i.e.~subtracting its mean and dividing by $\E[X_i]$, making use of the
positive homogeneity and cash invariance property of $\rho$.
If\ $\BX$ has non-zero excess return, i.e.~$\E[\BX] \neq  \Bx$, then the additional linear term ``$\Bphi$ times excess return'' arises, which enters the minimization of the risk of the surplus with respect to $\Bphi$ in a straight forward way.
Similarly, if $\BL$ has non-zero mean (claim size distributions are typically positive, the centered variable $\BL - \E[ \BL]$
is regarded instead.
The detailed justification of the simplifying assumption is transferred to the appendix.

The following lemma shows that the $\alpha$-quantile of the surplus $S(\Bphi)$ is well defined and states further preliminary results. We denote by $\one_A$ the indicator function of some set $A$; further $F_Y$, $\bar{F}_Y = 1-F_Y$, and $F_Y^{-1}$ denotes the cumulative distribution function, the tail function, and the quantile function of some scalar random variable $Y$, respectively.

\begin{lemma} \label{lem:PrelimResultsFromAssump}
{
Assume (\ref{eq:TechAssumpIntegr}) and (\ref{eq:TechAssumpLdensity}). Then for every $\Bphi\in\R_+^n$ and $\alpha\in(0,1)$
\begin{enumerate}[a)]
  \item $\Pa(S(\Bphi)\leq z) =\alpha$ has a unique solution $z = z_{\Bphi,\alpha}$, i.e.~the $\alpha$-quantile of $S(\Bphi)$ is well defined.
  \item    $\VaR_\alpha[S(\Bphi)] = -z_{\Bphi,\alpha}$ and $\ES_\alpha[S(\Bphi)] = -\alpha^{-1} \cdot\E[S(\Bphi)\cdot\one_{S(\Bphi) \leq z_{\Bphi,\alpha}} ]$. 
  \item $\Bphi\mapsto\rho[S(\Bphi)]$ is differentiable for both risk measures $\rho\in\{\VaR_\alpha, \ES_\alpha\}$.
      \label{lem:PrelimResults_SphiDiffbar}
  \item $\Bphi\mapsto\ES_\alpha[S(\Bphi)]$ is convex.
\end{enumerate}
}
\end{lemma}
We denote the quantile of $S(\Bphi)$ by $z_\Bphi$  omitting the subscript $\alpha$ when there is no confusion about the risk tolerance.
Part (a) and (c)
result basically from the implicit function theorem applied to $(z,\Bphi) \mapsto F_{S(\Bphi)}(z)$;
(b) is a consequence of the continuous distribution of $S(\Bphi)$,
and (d) follows from the convexity of the expected shortfall. The details of the proofs are transferred to the appendix.

\begin{remark}
{\rm
\begin{enumerate}[a)]
\item If $\BL$ has atoms, i.e.~does not admit a density, then the function $\Bphi \mapsto\VaR_\alpha[S(\Bphi)]$ might not be continuous
   but can have kinks at 
	the singular values of $\BL$.
\item Assumption (\ref{eq:TechAssumpLdensity}) can be relaxed; it suffices to assume that $\BL$ admits a strictly positive density in some  open set around $\{\Bl \in \R^n: \scp{\Bone}{\Bl} = F^{-1}_{\scp{\Bone}{\BL}}(1-\alpha)\}$.
\end{enumerate}
}
\end{remark}

We introduce some further notation:
for two scalar functions $a(t)$ and $b(t)$ we denote
\mbox{$a(t) = O\big(b(t)\big)$,}
$a(t) \sim b(t)$, or $a(t) = o\big(b(t)\big)$
as $ t \to t_0$,
if
$\limsup_{t\to t_0} |a(t) / b(t)| < \infty$, or
$\lim_{t\to t_0} a(t) / b(t) = 1$ or $=0$, 
respectively.
We call a vector $\BX$ of tradeable assets {\em admissible} if $X_i$ is strictly positive with unit mean and satisfies condition \eqref{eq:TechAssumpIntegr} for every $i=1,\dots,n$.

Recalling the well-known link between expected shortfall and value-at-risk
$\ES_\alpha[\cdot] = \alpha^{-1}\int_0^\alpha \VaR_\beta[\cdot] \, d\beta$,
we present a result concerning the integration with respect to the confidence level.
\begin{lemma}
\label{lem:QuantIntegr}
  Consider a real-valued random variable with strictly positive density $f$ which enables a continuous quantile function $F^{-1}$. Further consider a differentiable function $G:\R\to\R$ with $G(x)\to 0$ as $x\to\infty$. Then for every $\alpha \in (0,1)$
  $$\int_0^\alpha \frac{G'\circ F^{-1} (1-\beta)}{f\circ F^{-1} (1-\beta)} \, d\beta = - G\circ F^{-1} (1-\alpha) \, .$$
\end{lemma}
This result follows directly from the change of variable $\beta \to y:= F^{-1}(1-\beta)$,
which implies 
$d\beta = -f(y)dy$.

\section{Particular Value of $\phi$ (one-dimensional case)}
\label{sec:GenResults}

The results of this section only hold in the one-dimensional case,~i.e.~if $n=1$. We abandon in the sequel the subscript $i$ equal to one and refrain from matrix notation.
We identify a particular initial investment amount $\phi$ into the tradeable asset $X$ such that $\rho[S(\phi)]$ becomes fairly independent of the distribution of $X$.

To separate the distribution of the tradeable asset $X$ from the claim size $L$, we analyze the event  $\left\{ S(\phi) \leq -\phi\right\}$ for any $\phi \geq 0$ and derive the following equivalent events:
\begin{eqnarray}
  \left\{ S(\phi) \leq -\phi\right\}
  = \{ \phi\!\cdot\!(X-1)-X\!\cdot\! L \leq -\phi\}
  = \{ X\!\cdot\!(\phi-L) \leq 0\} 
  =\{\phi-L \leq 0\} = \{L\geq \phi\} \, ,
&&\label{eq:phiIsqL_SetEquiv}
\end{eqnarray}
where  the last but one equality follows from the strict positivity of $X$.
Hence we derive that $\Pa\left( S(\phi) \leq -\phi\right) = 1 - F_L(\phi)$.
As we are interested in the $\alpha$-quantile of $S(\phi)$, we need to choose $\phi =
\qL:= F_L^{-1}(1-\alpha)$,
which is well defined due to assumption (\ref{eq:TechAssumpLdensity}).
This implies  
$z_\qL = -\qL$ 
or, equivalently,
$\VaR_\alpha\left[ S(q)\right] = q$.

Also for the expected shortfall, $\phi = \qL$ is a special case: since
$\{S(\qL)\leq z_\qL\} 
= \{L\geq \qL\}$, which follows directly from (\ref{eq:phiIsqL_SetEquiv}),
we conclude
\begin{eqnarray}
-\alpha\cdot \ES_\alpha[S(\qL)] &=&  \E[S(\qL)\cdot \one_{S(\qL)\leq z_\qL}] 
=\E\left[ \big(\qL\cdot(X-1)-X\cdot L\big) \cdot\one_{L\geq \qL}\right] \label{eq:reformESSphi}\\
&=&  \qL\cdot \E[ X-1] \cdot \Pa(L\geq \qL) - \E[X] \cdot \E[L\cdot\one_{L\geq \qL} ] \nonumber\\
&=&   
  \E[-L\cdot\one_{-L\leq -\qL =F_{-L}^{-1}(\alpha) } ] =
-\alpha\cdot\ES_\alpha[-L] \, ,\nonumber
\end{eqnarray}
where the third equality follows from the independence of $X$ and $L$ and the forth equality from the unit mean of $X$.

Also the first derivative of the function $\phi \mapsto \rho[S(\phi)]$ shows special properties at $\phi = \qL$.
We summarize the findings in the following theorem together with all other results concerning the particular value for $\phi$.
\begin{theorem}
\label{thm:SpecialPoint_phi_ql}
Assume (\ref{eq:TechAssumpIntegr}) and (\ref{eq:TechAssumpLdensity}).
If $\qL := F_L^{-1}(1-\alpha) = \VaR_\alpha[-L]$ units are initially invested in $X$, i.e.~if $\phi=\qL$,  then
\begin{enumerate}[a)]
  \item $\rho[S(\qL)] = \rho[-L]$ for $\rho \in \{\VaR_\alpha,\ES_\alpha\}$.
  \item the differential of the risk of the surplus with respect to $\phi$ evaluated at $\phi = \qL$ reads
  \begin{eqnarray*}
        \big(\partial_\phi \, \rho[S(\phi)]\big)_{|_{\phi=\qL}} = \left\{
                 \begin{array}{lll}
                 (-1)\cdot \Big(\E[X^{-1}]^{-1}-1\Big) \geq 0 & \mbox{if} & \rho = \VaR_\alpha \, , \ \\ 	%
                   0 & \mbox{if} & \rho = \ES_\alpha \, .\\
                \end{array}
                \right.         \end{eqnarray*}
        and the above inequality becomes strict if $X$ is not constant.\footnote{Since in the expression for the value-at-risk the figure $(-1)$ appears four times in this formula, we propose the name \mbox{``4 x -1" formula}}
   \item the function $\phi\mapsto \ES_\alpha[S(\phi)]$ attains its global minimum value $\ES_\alpha[-L]$ at $\phi^* = \qL$. ($\phi^*$ is not necessarily unique.)
\end{enumerate}
\end{theorem}
Part (a) has already been shown above, the proof of (b) is transferred to the appendix, and (c) follows from (b) using the differentiability and convexity of  $\phi\mapsto \ES_\alpha[S(\phi)]$, see Lemma \ref{lem:PrelimResultsFromAssump}.

\begin{remark}
\label{rem:ThmSpecialPhi1dim}
{\rm
\begin{enumerate}[a)]
\item  
      The particular asset allocation $q$ is model-independent in the following sense: the risk $\rho[S(q)]$ becomes independent of the asset distribution for both risk measures value-at-risk and expected shortfall, as long as the asset is strictly positive.
\item The model-independent risk value at the particular asset value equals $\rho[-L]$ which coincides with the risk of the surplus if the volatility of $X$ collapse to zero and $X$ becomes constant (with value one).
\item The initial amount $\phi^*$ invested in $X$ that minimizes the risk $\rho[S(\phi)]$ is less than $\rho[-L]$ for both risk measures $\rho \in \{\VaR_\alpha,\ES_\alpha\}$.
   For $\VaR_\alpha$ this follows from part (b) of the theorem, for $\ES_\alpha$ the minimum is attained at $\phi^* = \VaR_\alpha[-L] < \ES_\alpha[-L]$. This phenomenon is due to the diversification  between $X$ and $L$. The probability of a synchronous realization of $X$ and $L$ beyond their respective $(1\!-\!\alpha)$-quantiles amounts to $\alpha^2 \ll \alpha$. Hence it makes sense to immunize against shocks in $X$ based on a claim size notional below $\rho[-L]$.
\item In the general multi-dimensional case we can not expect to find a particular asset allocation $\Bphi^*$ such that the risk of the surplus $\rho[S(\Bphi^*)]$ becomes independent of the distribution of the asset vector.
    The reason is that
    the separation of claims sizes from the tradeable assets 
does not work any more as in the univariate case. Similar to  \eqref{eq:phiIsqL_SetEquiv} we derive
    \mbox{$\left\{ S(\Bphi) \leq -\scp{\Bone}{\Bphi}\right\}
  = \{ \scp{\BX }{\Bphi-\BL} \leq 0\}$.} Due to the  scalar product structure the positivity of $\BX$ is not sufficient to deduce that $\Bphi-\BL$ is positive in all dimensions as in the univariate case.
\end{enumerate}
}
\end{remark}

\section{Expansion Results}
\label{sec:Perturb}

\subsection{Gram-Charlier-like expansion}

The classical Cornish-Fisher method \cite{CornishFisher} yields an expansion of the quantile of the surplus
based on its moments. 
These can be easily computed from 
\eqref{eq:DefSphiSimplif} in terms of the moments of $\BL$ and $\BX$ using
their independence.

Figure  \ref{fig:CornishFisher} compares the forth order Cornish-Fisher expansion with the true
value-at-risk profile of the surplus as a function of the asset allocation $\phi$ in the univariate case.
This Cornish-Fisher expansion fails to reproduce the relation
$\VaR_\alpha[S(q)] = q$ of Theorem \ref{thm:SpecialPoint_phi_ql}.(a) which holds independently of the distributions of $X$ and $L$.
The reason is that due to the product structure of the liability the third and higher moments of $S(\phi)$ differ considerably from those of the normal distribution.

\begin{figure}[h]
\centering
\includegraphics[width=1.0\textwidth,  angle = 0]{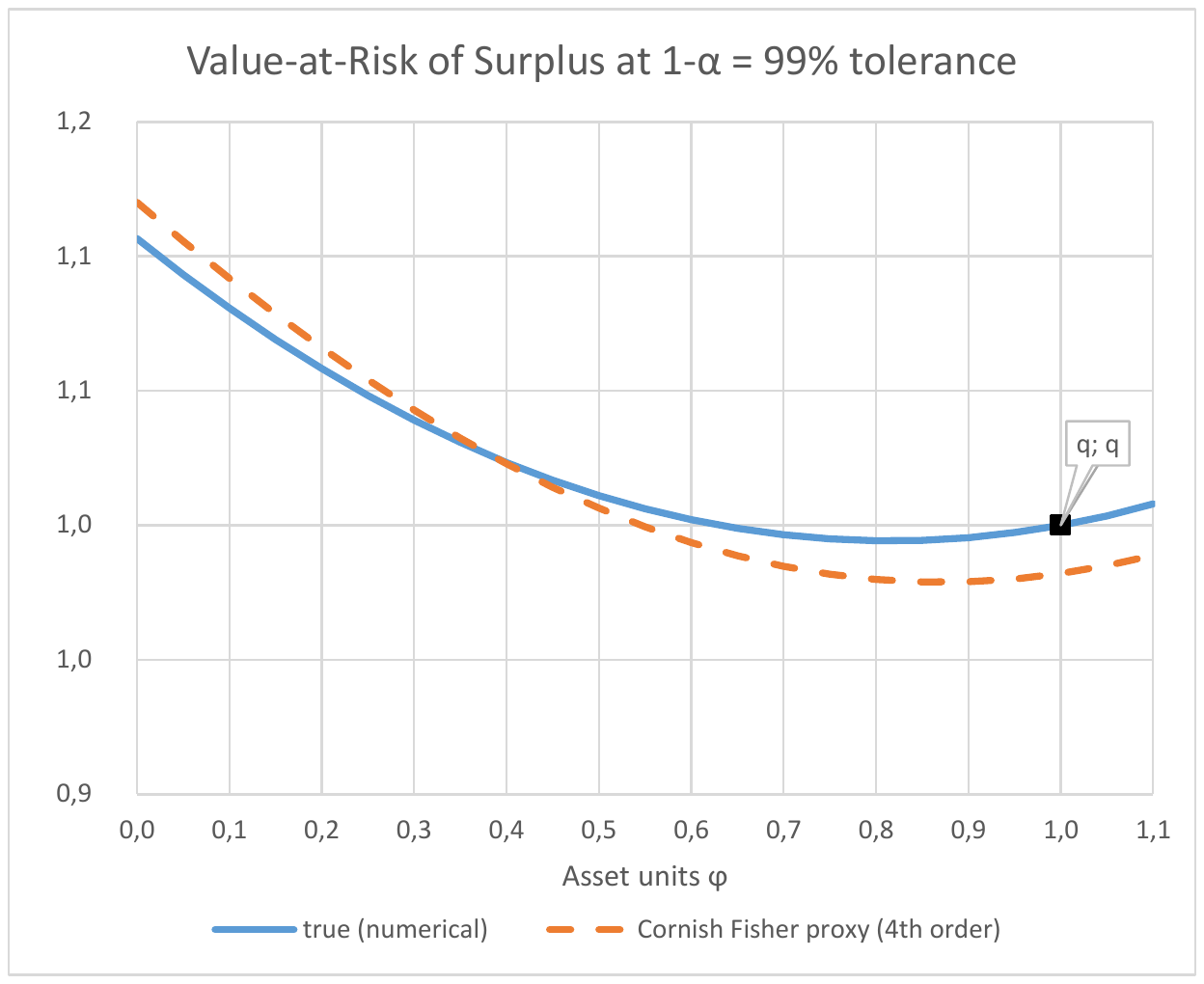}
\caption{\footnotesize True value and 4th order Cornish-Fisher approximation of the value-at-risk of the surplus as a function of the units $\phi$ of the financial asset $X$. The risk tolerance is set to $1-\alpha = 99\%$, the non-hedgeable component $L$ is normally distributed with $\sigma_L=0.43$ such that $q=$VaR$_\alpha$(-L)=1, and $X$ is log-normally distributed with log-normal volatility $\sigma = 0.25$. }
\label{fig:CornishFisher}
\end{figure}

%
We suggest an expansion 
that preserves the relation of Theorem \ref{thm:SpecialPoint_phi_ql}.(a).
To this aim we prove an expansion similar to the Gram-Charlier series \cite{Charlier} for the sum of two not necessarily independent
random variables. This expansion does not use the Gaussian distribution as base function but the distribution of one of the variables itself.

\begin{proposition}
\label{prop:ExpansSum2depRVs}
Consider two scalar random variables $Y_0$ and $Y_1$ such that
$Y_0+Y_1$ has a density which is differentiable for any order and the differentials are integrable.
Then
$$F_{Y_0 + Y_1}(z) = \Pa(Y_0 + Y_1 \leq z) = \sum_{r=0}^\infty \frac{1}{r!} \cdot (-D_z)^r\, \E\big[Y_1^r \cdot \one_{Y_0\leq z}\big] \, .$$
\end{proposition}
This theorem is proved by means of the Fourier transform; the details are transferred to the appendix.

\begin{remark}
{\rm
If $Y_0$ and $Y_1$ are independent, the expansion reads
$F_{Y_0 + Y_1} = \sum_{r=0}^\infty \frac{1}{r!} \cdot m_r(Y_1) \cdot (-D_z)^r \, F_{Y_0}$, where  $m_r(Y_1)$ denotes the r-th moment of $Y_1$.
This results is in line with classical Gram-Charlier series that are based on directly expanding the characteristic function instead of the cumulant generating function, refer to sec.~12 of \cite{JohnsonKotz}
}
\end{remark}


To apply Proposition \ref{prop:ExpansSum2depRVs} to
the surplus $S(\Bphi)= \scp{\BX - \Bone}{\Bphi} - \scp{\BX}{\BL} $
we rewrite it
in the form  $S(\Bphi)= Y_0 + Y_1$
with a purely non-hedgeable base function $Y_0 := - \scp{\Bone}{\BL}$
perturbed by a noise term $Y_1 := \scp{\BX-\Bone}{\Bphi - \BL}$ that depends linearly on the hedgeable asset.
%
An application of Proposition \ref{prop:ExpansSum2depRVs} leads
\begin{eqnarray}
\Pa(S(\Bphi) \leq z) &=& \Pa\big(-\scp{\Bone}{\BL} \leq z\big)
+  \sum_{i \geq 2} \frac{(-1)^i}{i!} \cdot D_z^i \, \E \Big[ \scp{\BX-\Bone}{\Bphi - \BL}^i \cdot \one_{-\scp{\Bone}{\BL} \leq z} \Big] \, .
\nonumber
\end{eqnarray}
The first order term vanishes since the terms involving $\BX$ and $\BL$ are independent and $\BX$ has unit mean.
Noting that $\scp{\BX-\Bone}{\Bphi - \BL}^i =
\sum_{j_1, \cdots, j_i = 1}^n \prod_{k = 1}^i
(X_{j_k}-1)\cdot (\phi_{j_k}-L_{j_k})$, we can again use
this independence 
to integrate the i-th order term with respect to the asset dimension to deduce
\begin{eqnarray}
\Pa(S(\Bphi) \leq z) &=& \bar{F}_{\scp{\Bone}{\BL}}[\Bphi-\BL](-z)
+ \sum_{i \geq 2} \; \frac{1}{i!} \; \cdot\sum_{j_1, \cdots, j_i = 1}^n   \bar{m}_{j_1, \cdots, j_i} \cdot D^i K_{j_1, \cdots, j_i}[\Bphi-\BL](-z) \, ,
\label{eq:ExpandCumulDistrSphiMomentsX}
\end{eqnarray}
where 
$K_{j_1, \cdots, j_i}[\Bphi-\BL]( y) := 
\E_\BL 
[\prod_{k = 1}^i (\phi_{j_k}-L_{j_k})  \cdot \one_{\scp{\Bone}{\BL} > y}]$
depends only on the claim size and
$\bar{m}_{j_1, \cdots, j_i}  := \E_\BX[\prod_{k = 1}^i (X_{j_k}-1)]$
represents the i-th multidimensional central moment of the tradeable assets; further
$\bar{F}_{\scp{\Bone}{\BL}}$ is the tail function of the random variable $\scp{\Bone}{\BL}$.
Note that the $(-1)^i$ terms have vanished since the terms $\one_{-\scp{\Bone}{\BL} \leq z}$ are now referenced
in the function $K_{j_1, \cdots, j_i}$ by the expression
$(\one_{\scp{\Bone}{\BL} \geq y})_{|_{y=-z}}$ and i-times differentiation reproduces these $(-1)^i$ terms.

\subsection{Second order expansion}

We have derived an expansion of the cumulative distribution of the surplus $S(\Bphi)$ in terms of the (multi-dimensional) moments of the tradeable assets $\BX$.
But what we need is an expansion of the $\alpha$-quantile $z = z(\Bphi)$ of 
$S(\Bphi)$
when the financial asset vector $\BX$ becomes more and more deterministic, i.e.~approaches the constant vector $\Bone$.

We denote by $\BSigma$ the
covariance matrix of the tradeable assets $\BX$,
i.e.~$\Sigma_{ij}= \E[(X_i - 1)\cdot (X_j - 1)]$.
We consider convergence of $\BX$ to $\Bone$ in quadratic norm, i.e.~$\|\BX-\Bone\|_2 := ( \E[\scp{\BX - \Bone}{\BX - \Bone}])^{1/2} \to 0$.
Note that
$\|\BX-\Bone\|_2^2 = \tr(\BSigma) = \|\BSigma\|_*$, where $\tr(\cdot)$ denotes the trace operator and $\|\cdot\|_*$ the nuclear norm.
Due to the equivalence of matrix norms there exists some constant $C>0$ such that for any vectors
$\Bu, \Bv 
\in \R^n$
$$|\scp{\Bu}{\BSigma \cdot \Bv}| \leq \|\BSigma\|_2 \| \cdot
\Bu\|_2\cdot \|\Bv\|_2
\leq C \cdot \|\BSigma\|_*\cdot \|\Bu\|\cdot \|\Bv\| =  C \cdot \|\BX-\Bone\|_2^2 \cdot\|\Bu\| \cdot\|\Bv\| \, .$$
This implies that for every $\Bu, \Bv \in \R^n$
\begin{equation}\
\label{eq:scpXSigXisOXmin1}
 \scp{\Bu}{\BSigma \cdot \Bv} = O(\|\BX-\Bone\|_2^2) \quad \mbox{as} \; \|\BX-\Bone\|_2\to 0\, .
\end{equation}
\begin{remark}
\label{rem:Xto1in2norm}
{\rm
\begin{enumerate}[a)]
  \item Relation \eqref{eq:scpXSigXisOXmin1} holds true independently of the particular convergence of $\BX \to\Bone$: for any family $(\BX_\sigma)_{\sigma>0}$ with
$\| \BX_\sigma  -\Bone\|_2 \sim \sigma$ as $\sigma\to 0$ and $\BX_\sigma$ admissible for every $\sigma>0$ we have
  $\scp{\Bu}{\BSigma_\sigma \cdot \Bv} = O(\sigma^2)$ as $\sigma\to 0$ where $\BSigma_\sigma$ denotes the covariance matrix of $\BX_\sigma$.
  \item The term $\scp{\Bu}{\BSigma_\sigma \cdot \Bv}$ can contain terms of higher order than $\sigma^2$ if some dimensions of $\BX$ converge faster to the constant than others, e.g. $\BX_\sigma = \big(1+\sigma \cdot (X_1-1), 1+\sigma^2 \cdot (X_2-1)\big)$ with some independent admissible $X_i$.
\end{enumerate}
}
\end{remark}
We choose an expansion of the $\alpha$-quantile $z$ of the surplus as $\sigma := \| \BX - \Bone \|_2 \to 0$
in the form
$$ \mbox{$z = z(\Bphi,\sigma) = \sum_{i = 0}^\infty z_i(\Bphi,\sigma)$ with
$z_i(\Bphi,\cdot) \sim \sigma^i$ as $\sigma \to 0$ for every $i\in\N_0$.}$$
When we insert the $\alpha$-quantile $z(\Bphi)$ into equation \eqref{eq:ExpandCumulDistrSphiMomentsX}, the left hand side equals $\alpha$ by definition of the quantile.
We then expand all $\sigma$-dependent terms of the right hand side of \eqref{eq:ExpandCumulDistrSphiMomentsX} in orders of $\sigma^i$. Note that only the moments of $\BX$ in the expansion \eqref{eq:ExpandCumulDistrSphiMomentsX} depend directly on $\sigma$;
all other terms depend only via the quantile $z$ on $\sigma$.
This enables us to evaluate sequentially the terms $z_i$ in increasing \mbox{order of $\sigma^i$.}

Let us start to expand the terms in equation \eqref{eq:ExpandCumulDistrSphiMomentsX} in orders of $\sigma^i$ as $\sigma \to 0$.
The first term of the right hand side of equation \eqref{eq:ExpandCumulDistrSphiMomentsX}
reads as $\sigma\to 0$
\begin{equation}
\label{eq:sigExpansFtail1L_}
\bar{F}_{\scp{\Bone}{\BL}}(-z)=
\bar{F}_{\scp{\Bone}{\BL}}(-z_0) -
f_{\scp{\Bone}{\BL}}(-z_0)\cdot (-z_1-z_2 - \dots)
- \mbox{$\frac{1}{2}$} f_{\scp{\Bone}{\BL}}'(-z_0)\cdot (-z_1- \dots)^2+\dots \, .
\end{equation}
We start to evaluate the zero and first order terms $z_0$ and $z_1$
of the quantile expansion.
Having \eqref{eq:sigExpansFtail1L_} in mind, relation \eqref{eq:ExpandCumulDistrSphiMomentsX}
reads for the $\alpha$-quantile
in first order approximation
\begin{equation*}
\alpha = \bar{F}_{\scp{\Bone}{\BL}}(-z_0 - z_1) + o(\sigma)=  \bar{F}_{\scp{\Bone}{\BL}}(-z_0) -  f_{\scp{\Bone}{\BL}}(-z_0)\cdot (-z_1) + o(\sigma) \, .
\end{equation*}
Collecting the zero order terms we obtain $1-\alpha = {F}_{\scp{\Bone}{\BL}}(-z_0)$. Denoting again
$\qL:= {F}_{\scp{\Bone}{\BL}}^{-1}(1-\alpha)$ we deduce that $-z_0 = q$.
Collecting the first order terms we obtain $ 0 = f_{\scp{\Bone}{\BL}}(q)\cdot z_1$. From the
positivity of the density $f_{\scp{\Bone}{\BL}}$ we conclude that $z_1 \equiv 0$.

Before we start the evaluation of the second order term $z_2$, we define some useful functionals:
\begin{eqnarray}
\BK(y) :=  \E_\BL \Big[ \BL\cdot \one_{\scp{\Bone}{\BL} > y} \Big]
\, , &&\quad
K_\BSigma[\BZ](y) :=  \E_\BL \Big[\scp{\BZ}{\BSigma \cdot \BZ}  \cdot \one_{\scp{\Bone}{\BL} > y} \Big]
 \, ,
\label{eq:DefK_EKone1Kgeqz}
\end{eqnarray}
for any $\R^n$-valued random variable $\BZ$.
%
This allows us to rewrite the second order term in the expansion \eqref{eq:ExpandCumulDistrSphiMomentsX}
as $\frac{1}{2} \cdot K_\BSigma[\Bphi - \BL]''(-z)$.
By equation \eqref{eq:scpXSigXisOXmin1} we know that $K_\BSigma[\Bphi - \BL](y) = O(\sigma^2)$ and hence also
$K_\BSigma[\Bphi - \BL]''(y) = O(\sigma^2)$ as $\sigma \to 0$ for every $y\in\R$.

To evaluate the second order term $z_2$ we collect
in the relation \eqref{eq:ExpandCumulDistrSphiMomentsX}
combined with the expansion \eqref{eq:sigExpansFtail1L_}
all terms $\sim \sigma^2$ as $\sigma \to 0$ and obtain
\begin{equation}
\label{eq:SecondOrdExpansStart}
0  = -  f_{\scp{\Bone}{\BL}}(-z_0)\cdot (-z_2) + \mbox{$\frac{1}{2}$} \cdot
K_\BSigma[\Bphi - \BL]''(-z_0)
+ o(\sigma^2) \, .
\end{equation}
The following theorem reformulates this second order expansion result for the value-at-risk of $S(\Bphi)$ and
derives the risk minimizing asset allocation.

\begin{theorem}\label{thm:PerturbVarMultiDim}
\begin{enumerate}[a)]
    \item
    Define $\qL:= \VaR_\alpha[-\scp{\Bone}{\BL}] = {F}_{\scp{\Bone}{\BL}}^{-1}(1-\alpha)$ and denote the covariance matrix of $\BX$ by $\BSigma$.
The expansion of
	$\VaR_\alpha[S(\Bphi)]$ up to second order in $\sigma:= \| \BX - \Bone\|_2 = \sqrt{\tr(\BSigma)}\to 0$  is given by
	\begin{eqnarray*}
	\VaR_\alpha[S(\phi)] & =& \qL + \mbox{$\frac{1}{2}$} \cdot f_{\scp{\Bone}{\BL}}(q)^{-1}
	\cdot K_\BSigma[\Bphi - \BL]''(q) + o(\sigma^2) \\
	& =& \qL - \mbox{$\frac{1}{2 f_{\scp{\Bone}{\BL}}(q)}$} \cdot
	\Big\{	\scp{\Bphi }{\BSigma  \!\cdot\!\Bphi } \!\cdot\! f_{\scp{\Bone}{\BL}}' (q)+ 2 \scp{\BSigma  \!\cdot\!\Bphi}{ \BK''(q) }
	- K_\BSigma[\BL]'' (q)	\Big\} + o(\sigma^2)\, .
	\end{eqnarray*}
\item
	If $f_{\scp{\Bone}{\BL}}'(\qL) \neq 0$ and $\BSigma$ is invertible, the minimum of the second order expansion of $\VaR_\alpha[S(\Bphi)]$ is attained at
	$\Bphi^* = -f_{\scp{\Bone}{\BL}}'(q)^{-1}\cdot \BK''(q) $
and equals
$$\VaR_\alpha[S(\Bphi^*)] = \qL + \mbox{$\frac{1}{2 f_{\scp{\Bone}{\BL}}(q)}$} \cdot
	\Big\{	f_{\scp{\Bone}{\BL}}' (q)^{-1} \!\cdot\! \scp{\BK''(\qL) }{\BSigma \cdot \BK''(\qL) } + K_\BSigma[\BL]'' (q)	\Big\} \, .$$
\end{enumerate}
\end{theorem}
\proof part a) follows from solving \eqref{eq:SecondOrdExpansStart} for $z_2$ and expressing $K_\BSigma[\Bphi \!- \!\BL]$ via the K-terms defined in \eqref{eq:DefK_EKone1Kgeqz}.
Differentiating the second equation of part a) with respect to $\Bphi$, setting it to zero, and multiplying from the left by
$f_{\scp{\Bone}{\BL}}(q) \cdot \BSigma^{-1}$ proves
the first assertion of
part b). Inserting this into the second equation of part a) yields the second assertion.
\qed

\begin{remark}
{\rm
The investment amount  $\Bphi^*$ in the tradeable assets that minimizes the second order expansion of $\VaR_\alpha[S(\Bphi)]$
(when the asset volatility tends to zero)
 is completely independent of the asset distribution.
Only the value-at-risk of the surplus at the optimal asset allocation $\Bphi^*$ depends on the assets via $\BSigma$.
}
\end{remark}

We now turn to the expected shortfall of the surplus which can be characterized in terms of the value-at risk by
$\ES_\alpha[S(\Bphi)] = \alpha^{-1}\int_0^\alpha \VaR_\beta[S(\Bphi)] \, d\beta$. Its expansion
is an immediate consequence of Lemma \ref{lem:QuantIntegr} when setting
$G := K_\BSigma[\Bphi \!-\! \BL]'$.
\begin{corollary}
\label{corr:PerturbESMultiDim}
\begin{enumerate}[a)]
\item
The expansion of $\ES_\alpha[S(\Bphi)]$ up to second order in $\sigma = \|\BX-\Bone\|_2\to 0$
\mbox{is given by}
\begin{eqnarray*}
\ES_\alpha[S(\phi)] & =& \ES_\alpha[-\scp{\Bone}{\BL}]
- \mbox{$\frac{1}{2\alpha}$} \cdot K_\BSigma[\Bphi - \BL]'(q) + o(\sigma^2)\\
& =& \ES_\alpha[-\scp{\Bone}{\BL}]
  + \mbox{$\frac{1}{2\alpha}$} \Big\{
		\scp{\Bphi }{\BSigma  \!\cdot\!\Bphi } \!\cdot\! f_{\scp{\Bone}{\BL}} (q)+ 2 \scp{\BSigma  \!\cdot\!\Bphi}{ \BK'(q) }- K_\BSigma[\BL]'
		 (q)\Big\} + o(\sigma^2)\, .
\end{eqnarray*}
  \item
If $\BSigma$ is invertible, the minimum of the second order expansion of $\ES_\alpha[S(\Bphi)]$ is attained at
\mbox{$\Bphi^* = -f_{\scp{\Bone}{\BL}}(q)^{-1}\cdot \BK'(q) $}
and equals
$$\ES_\alpha[S(\Bphi^*)] = \ES_\alpha[-\scp{\Bone}{\BL}]
  - \mbox{$\frac{1}{2\alpha}$} \Big\{	f_{\scp{\Bone}{\BL}} (q)^{-1} \!\cdot\! \scp{\BK'(\qL) }{\BSigma \cdot \BK'(\qL) } + K_\BSigma[\BL]' (q)	\Big\} \, .$$
\end{enumerate}
\end{corollary}

We analyze the \emph{total optimal investment amount} $\Phi^*:= \sum_i \phi^*_i = \scp{\Bone}{\Bphi^*}$ in all tradeable assets defined as 
the sum of the optimal investment amounts $\phi^*_i$ in the tradeable assets $X_i$ that minimize
the second order expansion of $\rho[S(\Bphi)]$.
We establish a link to the \textit{associated single-asset case} that is characterized as follows: there is only one  tradeable asset $X_0$, i.e.~$X_i = X_0$ for every $i=1, \dots, n$, and the surplus  reads  $S_0(\phi_0) = \phi_0 \cdot (X_0-1) - X_0 \cdot \scp{\Bone}{\BL}$, where $\phi_0 >0$ is the
investment amount into this single asset. We denote by $\phi_0^*$ the optimal investment amount that
minimizes the second order expansion of  $\rho[S_0(\phi_0)]$ in the associated single asset case.

%
\begin{theorem}
\label{corr:OptPhiAlloc}
In second order approximation of  $\rho[S(\Bphi)]$ according to Theorem \ref{thm:PerturbVarMultiDim} the
total optimal investment amount $\Phi^*$ satisfies:
\begin{enumerate}[a)]
\item $\Phi^* = \qL + {f_{\scp{\Bone}{\BL}}(q)}/{f_{\scp{\Bone}{\BL}}'(q)}$ if $\rho = \VaR_\alpha$, and
$\Phi^*=\qL$ if $\rho = \ES_\alpha$.
\item $\Phi^* = \phi_0^*$ for $\rho \in \{\VaR_\alpha, \ES_\alpha\}$, i.e.~the total optimal investment amount coincides with the optimal investment amount in the associated single-asset case.
\end{enumerate}
\end{theorem}
\proof 
we denote by $K_{\scp{\Bone}{\BL} }(z) := \E[\scp{\Bone}{\BL}\cdot\one_{\scp{\Bone}{\BL} >z}]= \int_q^\infty  t\cdot f_{\scp{\Bone}{\BL}} (t) \, dt$.
Observe that $\Phi^*=\scp{\Bone}{\Bphi^*}=- K_{\scp{\Bone}{\BL} }''(q) / f_{\scp{\Bone}{\BL}}'(q)$ if $\rho=\VaR_\alpha$
by Theorem \ref{thm:PerturbVarMultiDim} and  
$= - K_{\scp{\Bone}{\BL} }'(q) / f_{\scp{\Bone}{\BL}}(q)$
if $\rho=\ES_\alpha$ by Corollary  \ref{corr:PerturbESMultiDim}.
Further  note that
$K_{\scp{\Bone}{\BL} }'(q) = -q \cdot f_{\scp{\Bone}{\BL}} (q)$ and
$K_{\scp{\Bone}{\BL} }''(q) = -q \cdot f_{\scp{\Bone}{\BL}}' (q)-f_{\scp{\Bone}{\BL}}(q) $, which proves part a).
As a) also holds in the one-dimensional case, part b) follows by inspection of the formula in a) in the one-dimensional associated single-asset case. \qed
%

Hence $\Bphi^*$ can be interpreted as an allocation of $\phi_0^*$ in the sense that $\sum_i \phi_i^* = \phi_0^*$.
We investigate the impact of the multivariate
claim size distribution on this allocation:
if a particular claim size $L_i$ is more volatile and only weakly correlated
to the other
claim sizes $L_j$, $j\neq i$, then a material amount in the asset $X_i$ should show up in the risk-minimal asset allocation $\Bphi^*$.
If the claim sizes are multivariate normally distributed we obtain the following result, the proof of which is transferred to the appendix.
\begin{theorem}
\label{thm:AllocPhiOptLmultiNormal}
Assume that the claim sizes $\BL \sim \Ncal(\Bzero, \BSigmaL)$ follow a multivariate normal distribution with covariance matrix $\BSigmaL$.
\begin{enumerate}
  \item
Then for $\rho \in \{\VaR_\alpha, \ES_\alpha\}$ the investments $\phi_i^*$ in the tradeable assets $X_i$ that minimize
 $\rho[S(\Bphi)]$ expanded up to second order in the asset volatility $\sigma = \|\BX-\Bone\|_2\to 0$ follow the \emph{covariance allocation principle} with respect to  $\BL$, i.e.
$$\phi_i^* = \frac{\Sigma^\BL_{ii} + \sum_{j \neq i} \Sigma^\BL_{ij}}{\scp{\Bone}{\BSigmaL  \!\cdot\!\Bone}} \cdot \phi_0^* \qquad (i = 1, \dots, n) \, ,$$
where $\phi_0^*$ is the risk-minimal investment in the associated single-asset case according to \mbox{Theorem \ref{corr:OptPhiAlloc}}
and $\scp{\Bone}{\BSigmaL  \!\cdot\!\Bone}$ 
is the total variance of $\sum_i L_i$.
\item The minimum of the risk of the surplus $\rho[S(\Bphi^*)]$ in second order approximation for $\rho \in \{\VaR_\alpha, \ES_\alpha\}$ equals
\begin{eqnarray*}
  \VaR_\alpha[S(\Bphi^*)] &=& q + \frac{(\ln f_{\scp{\Bone}{\BL}})'(q)}{2}  \cdot \left\{ \left( 1+ \frac{(\ln f_{\scp{\Bone}{\BL}})'(q)^{-2}}{\scp{\Bone}{\BSigmaL\cdot \Bone }} \right)\cdot
   \frac{\scp{\Bone}{\BSigmaL\cdot\BSigma\cdot\BSigmaL \cdot \Bone }}{\scp{\Bone}{\BSigmaL\cdot \Bone }} - \tr(\BSigma \cdot\BSigmaL)
   \right\} \, ,\\
  \ES_\alpha[S(\Bphi^*)] &=& \ES_\alpha[-\scp{\Bone}{\BL}]
  - \frac{f_{\scp{\Bone}{\BL}}(q)}{2\alpha} \cdot \left\{	 \frac{\scp{\Bone}{\BSigmaL\cdot\BSigma\cdot\BSigmaL \cdot \Bone }}{\scp{\Bone}{\BSigmaL\cdot \Bone }}
-\tr(\BSigma \cdot\BSigmaL) \right\} \, .
  \end{eqnarray*}
\end{enumerate}
\end{theorem}

Theorem \ref{thm:PerturbVarMultiDim} and Corollary  \ref{corr:PerturbESMultiDim}
describe the expansion results in terms of derivatives of the K-terms defined in \eqref{eq:DefK_EKone1Kgeqz}.
In order to calculate these terms explicitly a rotation in the state space of $\BL$ proofs useful: let $\BD \in SO(n)$ be a rotation matrix in the $n$-dimensional special orthogonal group\footnote{I.e.~$\BD$ has unit determinate and  pairwise orthogonal columns with unit $l_2$-norm}, such that the first column of $\BD$
is parallel to the $\Bone$ vector.
%
The rotation matrix can be written
$\BD = \big( n^{-1/2}  \!\cdot\! \Bone \big| \BoneT \big)$, where $\BoneT$ is a $n\times(n\!-\!1)$ matrix of  orthogonal coordinates that span the hyperplane orthogonal to the vector $\Bone$.
In two and three dimensions the rotation matrix $\BD$ reads
$$\BD_{(n=2)} = \mbox{$\frac{1}{\sqrt{2}}$} \cdot\begin{pmatrix}
1 & -1 \\
1 & 1
\end{pmatrix} \, , \quad
\BD_{(n=3)} = \mbox{$\frac{1}{\sqrt{6}}$}  \cdot\begin{pmatrix}
  \sqrt{2} & 1 & -\sqrt{3} \\
  \sqrt{2} & 1 &  \sqrt{3} \\
  \sqrt{2} &-2 & 0 \end{pmatrix} \, .$$
Rewriting $\BK(y) =  \int_{\{\Bl\in\R^n: \scp{\Bone}{\Bl} > y\}} \Bl \cdot f_\BL(\Bl) \, d\Bl$
we apply  the change in variable $\Blambda := \BD'\Bl$ (implying $\Bl = \BD\Blambda$), which yields
\begin{eqnarray}
\label{eq:RewrKzTrafoVariab}
\BK(y)
= \int_{\{\Blambda\in\R^n: \scp{\Bone}{\BD\Blambda} > y\}} \BD\Blambda \cdot  f_\BL(\BD\Blambda) \, d\Blambda
= \int_{\R^{n-1}} \int_{y/\sqrt{n}}^\infty  \Big( \frac{\lambda_1}{\sqrt{n}} \cdot \Bone + \BoneT\bar{\Blambda} \Big) \cdot g(\lambda_1, \bar{\Blambda}) \; d\lambda_1 \, d\bar{\Blambda} \, ,&&\end{eqnarray}
where
$g(\Blambda) : = f_\BL(\BD\Blambda)$ denotes the rotated density. The last equation follows from the observation that
$\scp{\Bone }{\BD\Blambda } = \scp{\Bone }{n^{-1/2} \cdot \lambda_1 \cdot \Bone + \BoneT \cdot \bar{\Blambda}}
= \sqrt{n} \cdot \lambda_1$.
A similar expression can be derived for
$K[\BL](y)$.
The following result reformulates the derivatives of the K-terms
accordingly.

\begin{theorem}
\label{thm:RepresK_Terms}
Defining the expressions
$$\Bh(y) := \mbox{$\frac{1}{\sqrt{n}}$} \int_{\R^{n-1}}  \bar{\Blambda} \cdot g\big(\mbox{$\frac{y}{\sqrt{n}}$} , \bar{\Blambda}\big)  \, d\bar{\Blambda} \, , \quad
h_2(y)
:= \mbox{$\frac{1}{\sqrt{n}}$} \int_{\R^{n-1}}  \scp{\bar{\Blambda} }{\BoneT'\cdot\BSigma\cdot\BoneT\cdot\bar{\Blambda}}
\cdot g\big(\mbox{$\frac{y}{\sqrt{n}}$} , \bar{\Blambda}\big)  \, d\bar{\Blambda} \, ,$$
the first and second derivative of the K-terms defined in \eqref{eq:DefK_EKone1Kgeqz} reads
\begin{enumerate}[a)]
\item $\BK'(y) = - \mbox{$\frac{y}{{n}}$} \cdot f_{\scp{\Bone}{\BL}}(y) \cdot \Bone - \BoneT \cdot \Bh(y)$,
\item $K_\BSigma[\BL]' (y) = - \mbox{$\frac{y^2}{{n^2}}$} \cdot \scp{\Bone}{\BSigma \cdot\Bone} \cdot f_{\scp{\Bone}{\BL}}(y)
- \mbox{$\frac{2y}{{n}}$} \cdot \scp{\BoneT' \cdot\BSigma \cdot\Bone}{ \Bh(y)} - h_2(y) $,
\item $\BK''(y) = - \frac{1}{{n}} \cdot\big( f_{\scp{\Bone}{\BL}}(y) + y\cdot f_{\scp{\Bone}{\BL}}'(y)\big) \cdot \Bone - \BoneT \cdot \Bh'(y)$,
\item $K_\BSigma[\BL]''(y) = - \mbox{$\frac{y}{{n^2}}$} \cdot  \scp{\Bone}{\BSigma  \!\cdot\!\Bone} \cdot\big( 2 f_{\scp{\Bone}{\BL}}(y) + y \!\cdot\! f_{\scp{\Bone}{\BL}}'(y) \big)
- \mbox{$\frac{2}{{n}}$} \cdot \big\langle \BoneT' \!\cdot\!\BSigma  \!\cdot\!\Bone,  \Bh(y)+y \!\cdot\!\Bh'(y)\big\rangle
- h_2'(y) $.
\end{enumerate}
The minimum values of part (b) of Theorem \ref{thm:PerturbVarMultiDim} and Corollary \ref{corr:PerturbESMultiDim} read
\begin{enumerate}[a)]
\setcounter{enumi}{4}
\item $\VaR_\alpha[S(\Bphi^*)] = \qL + \mbox{$\frac{1}{2 f_{\scp{\Bone}{\BL}} (q)}$} \cdot
	\Big\{
\frac{f_{\scp{\Bone}{\BL}} (q)^2}{n^2 f_{\scp{\Bone}{\BL}}' (q)}
    \cdot \scp{\Bone}{\BSigma \cdot\Bone}
+ \frac{1}{f_{\scp{\Bone}{\BL}}'(q)}  \cdot
\Big\langle\Bh'(q), \, \BoneT' \!\cdot\!\BSigma \!\cdot\!\BoneT\!\cdot\! \Bh'(q) \Big\rangle$\\
$+ \frac{2}{n} \cdot
 \Big\langle \ln(f_{\scp{\Bone}{\BL}})'(q) \!\cdot\!\Bh'(q) - \Bh(q), \, \BoneT' \!\cdot\!\BSigma\!\cdot\!\Bone \Big\rangle
-  h_2'(y)\Big\} $.
\item $\ES_\alpha[S(\Bphi^*)] = \ES_\alpha[-\scp{\Bone}{\BL}]
  - \mbox{$\frac{1}{2\alpha}$} \Big\{ f_{\scp{\Bone}{\BL}} (q)^{-1} \!\cdot\!
\Big\langle\Bh(q), \, \BoneT' \!\cdot\!\BSigma \!\cdot\!\BoneT\!\cdot\! \Bh(q) \Big\rangle - h_2(y)\Big\} $.
\end{enumerate}
\end{theorem}

\proof
 the relation
$\frac{1}{\sqrt{n}} \int_{\R^{n-1}}
g\big(\mbox{$\frac{y}{\sqrt{n}}$} , \bar{\Blambda}\big)  \, d\bar{\Blambda}=
 D_y \int_{\{\Bl\in\R^n:\scp{\Bone}{\Bl}>y\}} d\Bl = f_{\scp{\Bone}{\BL}}(y)$
 is derived analogously to \eqref{eq:RewrKzTrafoVariab}.
Part a) follows from differentiating \eqref{eq:RewrKzTrafoVariab}
and applying this relation.
Part b) follows analog to a); c) and d) is obtained by differentiating a) and b) again.
Part e) and f) are obtained by inserting part a) to d) into the corresponding expressions of
Theorem \ref{thm:PerturbVarMultiDim} and Corollary \ref{corr:PerturbESMultiDim}, respectively. \qed


\subsection{Higher order expansion}

Deriving the third and higher order expansion terms is in principle straight forward but tedious, since the higher order expansion results are not any more independent of the specific convergence of the asset vector $\BX$ to the constant $\Bone$, refer to  Remark \ref{rem:Xto1in2norm}(a).
Let us choose a family $(\BX_\sigma)_{\sigma>0}$ of admissible asset vectors with $\| \BX_\sigma  -\Bone\|_2 \sim \sigma$ as $\sigma\to 0$ as in Remark \ref{rem:Xto1in2norm}(a).
In order to expand the Gram-Charlier-like formula \eqref{eq:ExpandCumulDistrSphiMomentsX} to third or higher order in $\sigma$  as $\sigma \to 0$ we need to expand the i-th central moments
$\bar{m}_{j_1, \cdots, j_i}(\sigma)  := \E_\BX[\prod_{k = 1}^i (X_{\sigma,j_k}-1)]$ in terms of $\sigma$ as follows
\begin{equation}\label{eq:ExpandMultiDimMoments}
\bar{m}_{j_1, \cdots, j_i}(\sigma) =    \bar{m}_{j_1, \cdots, j_i}^{(0)} \cdot\sigma^i
                                        + \bar{m}_{j_1, \cdots, j_i}^{(1)} \cdot\sigma^{i+1} + \dots
                                        + \bar{m}_{j_1, \cdots, j_i}^{(k)} \cdot\sigma^{i+k} + o(\sigma^{i+k}) \quad \mbox{as} \; \sigma \to 0\, .
\end{equation}
Recall that also for the second moments third and higher order terms can appear, refer to Remark \ref{rem:Xto1in2norm}(b).

Extending  equation \eqref{eq:SecondOrdExpansStart}, from which we derived  the second order terms, up to third order, we derive from \eqref{eq:ExpandCumulDistrSphiMomentsX} using \eqref{eq:sigExpansFtail1L_}
\begin{eqnarray*}
0  &=& -  f_{\scp{\Bone}{\BL}}(-z_0)\cdot (-z_2- z_3)
 + \mbox{$\frac{1}{2}$} \cdot \sum_{i, j =1}^n   \Big(\bar{m}_{i,j}^{(0)} \cdot\sigma^2 + \bar{m}_{i,j}^{(1)}\cdot \sigma^3 \Big) \cdot K_{i,j}[\Bphi-\BL]''(-z_0)\\
 && + \mbox{$\frac{1}{6}$} \cdot  \sum_{i, j,k =1}^n   \bar{m}_{i,j,k}^{(0)} \cdot\sigma^3 \cdot K_{i,j,k}[\Bphi-\BL]'''(-z_0) \, .
\end{eqnarray*}
Solving for the third order term $z_3$ we obtain the following result.
\begin{theorem}\label{thm:3rdOrderMultiDim}
Let us choose a family $(\BX_\sigma)_{\sigma>0}$ of admissible asset vectors with $\| \BX_\sigma  -\Bone\|_2 \sim \sigma$ as $\sigma\to 0$
and consider the expansion of the higher order moments as in \eqref{eq:ExpandMultiDimMoments}. Then the third order expansion of the value-at-risk of the surplus in $\sigma$ reads
	\begin{eqnarray*}
	\VaR_\alpha[S(\phi)] & =& \qL + \frac{\sigma^2}{2 \cdot f_{\scp{\Bone}{\BL}}(q)}
	\cdot K_{\BSigma^{(0)}}[\Bphi - \BL]''(q) \\
    && + \frac{\sigma^3}{6 \cdot f_{\scp{\Bone}{\BL}}(q)} \cdot \Big\{ 3 \cdot K_{\BSigma^{(1)}}[\Bphi - \BL]''(q)
    + \sum_{i, j,k =1}^n   \bar{m}_{i,j,k}^{(0)}  \cdot K_{i,j,k}[\Bphi-\BL]'''(q)  \Big\}    + o(\sigma^3) \, ,
	\end{eqnarray*}
where $\BSigma^{(k)} :=(\bar{m}_{i,j}^{(k)})_{ij}$ denotes the matrices of the expansion of the second order moments according to \eqref{eq:ExpandMultiDimMoments}
and the term $K_{i,j,k}$ is defined in \eqref{eq:ExpandCumulDistrSphiMomentsX}.
\end{theorem}

In the sequel we
demonstrate the effects of particular converging families of asset distributions that are important in practice
and  derive the forth order terms.
 Due to the increased complexity, we restrict to the \textbf{one-dimensional case}, i.e.~$n=1$.

The expansion
\eqref{eq:ExpandCumulDistrSphiMomentsX} of the cumulative distribution of the surplus then reads
in the one-dimensional case
\begin{eqnarray}
\Pa(S(\Bphi) \leq z) &=&
\bar{F}_{L}(-z) + \sum_{i \geq 2} \frac{\bar{m}_i}{i!} \cdot D^i \II_i(-z) \, , \quad
\mbox{where} \quad \II_i(y) :=
\int_y^\infty (\phi-\ell)^i  \cdot f_L(\ell) \, d\ell\, ,
\label{eq:n1CumDistSurplExp}
\end{eqnarray}
and $\bar{m}_i$ 
denotes the i-th central moment of the tradeable asset $X$.

%
A straight forward way to construct a family of admissible assets converging to the constant 1 is to scale a fixed asset variable $X$ by its normal volatility.
In financial application also the log-normal volatility is of high importance.
Hence we
introduce for a fixed tradeable asset $X$ the following two families of admissible assets $(X_\sigma)_{\sigma \geq 0}$ indexed by the normal as well as log-normal volatility:
we set $X_{\sigma_N} := 1 + \sigma_N Y$ in the normal case and $X_{\sigma_{lN}} := e^{\sigma_{lN}Y} / M(\sigma_{lN})$
in the log-normal case, where $Y$ denotes the
centered and normalized version of $X$ or $\ln X$, respectively,\footnote{
Normal case: $Y = (X-1)/\sqrt{\Var[X]}$, log-normal case: $Y = (\ln(Y) - \E[\ln Y]) /\sqrt{\Var[\ln X]}$. }
and $M(\sigma) := \E[e^{\sigma Y}]$ is the moment generating function of $Y$.
Note that the standard deviation of $X_{\sigma_{N}}$ or $\ln X_{\sigma_{lN}}$
equals ${\sigma_{N}}$ or ${\sigma_{lN}}$, respectively.
Further $X_{\sigma^*}$ coincides with the original tradeable asset $X$
if $\sigma^* = \sqrt{\Var[X]}$ in the normal and $=\sqrt{\Var[\ln X]}$ in the log-normal case.
Moreover, $X_{\sigma_{N}}$ and $\ln X_{\sigma_{lN}}$ keep the unit mean property due to the normalization.
Hence $X_{\sigma}$ is admissible for every $\sigma >0$ in the normal as well as in the log-normal case.

The central moments $\bar{m}_i = \bar{m}_i(\sigma):=\E[(X_\sigma-1)^i]$ of $X_\sigma$ for $\sigma \in \{\sigma_N, \sigma_{lN}\}$ show the following expansions in terms of the normal and log-normal asset volatility: denote
by $\mu_i := \E[Y^i]$ the i-th moment of $Y$, which coincides with the i-th centered and normalized moment of $X$ or $\ln X$, respectively.
In the normal case the expansion of $\bar{m}_i$ is trivially given by $\bar{m}_i = \sigma_N^i \cdot \mu_i$, whereas in the log-normal case the expansion of $\bar{m}_i$ up to forth order in $\sigma_{lN}$ reads
\begin{eqnarray}
\bar{m}_2 &=& \sigma_{lN}^2 + \mu_3 \cdot\sigma_{lN}^3 +\left(\mbox{$\frac{7}{12}$}\mu_4 - \mbox{$\frac{5}{4}$}\right)\cdot\sigma_{lN}^4 +o(\sigma_{lN}^4) \, , \nonumber\\
\bar{m}_3 &=& \mu_3 \cdot\sigma_{lN}^3  +\mbox{$\frac{3}{2}$}(\mu_4-1) \cdot\sigma_{lN}^4 +o(\sigma_{lN}^4) \, , 
\label{eq:MomentsXlognormExpans}\\
\bar{m}_4 &=& \mu_4 \cdot\sigma_{lN}^4 +o(\sigma_{lN}^4) \, .\nonumber
\end{eqnarray}

We summarize the results for the fourth order expansion of the $\VaR_\alpha[S(\phi)]$ in the following theorem.
The proof is transferred to the appendix together with proof of \eqref{eq:MomentsXlognormExpans}.
We denote by $id$ the identity function.

\begin{theorem}\label{thm:PerturbVar}
Consider the one-dimensional case, i.e.~$n=1$.
\begin{enumerate}[a)]
\item	The expansion of  $\VaR_\alpha[S(\phi)]$ in the log-normal volatility $\sigma_{lN}$
	of the financial asset $X$ up to fourth order as $\sigma_{lN}\to 0$ is given by
	\begin{eqnarray*}
		\lefteqn{\VaR_\alpha[S(\phi)]  = \qL
			-  \frac{1}{f_L(q)} \cdot \Bigg\{  \frac{\sigma_{lN}^2}{2} \cdot \big[(\phi-id)^2\!\cdot\! f_L\big]'(q)
			+ \frac{\sigma_{lN}^3 \!\cdot\!\mu_3}{6} \cdot\big[(\phi-id)^3 \!\cdot\!f_L'\big]'(q) } \\
&& + \frac{\sigma_{lN}^4}{24}\cdot\bigg[  \mu_4\!\cdot\!\big[(\phi-id)^4 \!\cdot\!f_L'\big]'
			  - 3\cdot\frac{{\big((\phi-id)^2 \!\cdot\!f_L\big)'\,}^2}{f_L}
				+(2\mu_4- 6) \cdot (\phi-id)^3\!\cdot\! f_L'\\
&&+ \left(\mu_4 +3\right) \cdot (\phi-id)^2 \!\cdot\!f_L	\bigg]'(q) \Bigg\} +   o(\sigma_{lN}^4) \, ,
	\end{eqnarray*}
 where $\mu_3$ and $\mu_4$ denote the third and forth centered normalized moments of $\ln X$, respectively.
\item If $\mu_3\cdot f_L''(q) \neq 0$, the  expansion of $\VaR_\alpha[S(\phi)]$ in (a) up to third order attains its local minimum at
		$$\phi^* = \qL + f_L''(q)^{-1} 
		\cdot \left( (1-\delta) \cdot f_L'(q) - \sqrt{(1-\delta)^2 \cdot f_L'(q)^2 + 2\cdot\delta\cdot f_L''(q)\cdot f_L(q) } \right) \, ,
		\quad \delta:= \mbox{$\frac{1}{\sigma\cdot\mu_3}$.}$$
If $\mu_3\cdot f_L''(q) = 0$ but $f_L'(\qL) \neq 0$, the minimum  is attained at
	$\phi^* = \qL + f_L(\qL) / f_L'(\qL)$.
\end{enumerate}
\end{theorem}

\begin{remark}
{\rm
  \begin{enumerate}[a)]
  \item The expansion of $\VaR_\alpha[S(\phi)]$ only involves local properties of $L$ around its $(1\!-\!\alpha)$-quantile, i.e.~(higher order)  derivatives of $f_L$ at $\qL$.
  \item If the skew of $\ln(X)$ vanishes and $L$ is normally distributed with volatility $\sigma_L$, then
    $\qL = \sigma_L \cdot u_{1-\alpha}$ where $u_{1-\alpha}$ denotes  the $(1\!-\!\alpha)$-quantile of the standard normal distribution. Hence $f_L'(\qL) / f_L(\qL) = -\qL/\sigma_L^2 = - u_{1-\alpha}/\sigma_L$. Part (b) of the theorem implies that  $\phi^*/\qL = 1-u_{1-\alpha}^{-2}$, which amounts to $0.815$ or $0.849$ for the risk tolerance $1\!-\!\alpha = 0.99$ (Basel II) or	$= 0.995$ (Solvency II), respectively. This means that the total Solvency II capital requirement of an insurance undertaking (when evaluated via a fully stochastic model) is minimized, if in addition to the expected claim size also $84.9\%$ of the non-hedgeable risk component, i.e.~the $99.5\%$-quantile of the centered claim size $L$, is initially invested in $X$.
\item The presence of a negative log-normal asset skew 
    (the usual case in practical applications) shifts the optimal asset allocation $\phi^*$ nearer to the $1\!-\!\alpha$ quantile $q$ of $L$, refer to Figures \ref{fig:VaRES_vs_phi_highAssetVol} and \ref{fig:phi*_vs_sigma}. The reason is that the diversification effect that reduces the risk minimal asset allocation $\phi^*$ to a value lower than $q$, refer to Remark \ref{rem:ThmSpecialPhi1dim}(c), is less pronounced if $\ln X$ is negatively skewed. Vice versa for a positive log-normal skew of $X$.
\end{enumerate}
}
\end{remark}

Repeating the proof of the expansion in the above theorem
using the normal instead of the log-normal asset volatility 
gives the following results.
\begin{corollary}
In the one-dimensional case, the expansion of  $\VaR_\alpha[S(\phi)]$ in the \emph{normal} asset volatility $\sigma_N$ 
up to forth order as $\sigma_N \to 0$ is given by
	\begin{eqnarray*}
		\VaR_\alpha[S(\phi)] & =& \qL
		-  \frac{1}{f_L(q)} \cdot \Bigg\{ \frac{\sigma_N^2}{2}\cdot \big[(\phi-id)^2 \!\cdot\!f_L\big]'(q)
		+ \frac{\sigma_N^3\!\cdot\! \mu_3}{6}\cdot \big[(\phi-id)^3 \!\cdot\!f_L\big]''(q)   \\
&& +\frac{\sigma_{N}^4}{24}\cdot\bigg[  \mu_4\cdot \big((\phi-id)^4 \!\cdot\!f_L\big)''
			  - 3\cdot\frac{{\big((\phi-id)^2\!\cdot\! f_L\big)'\,}^2}{f_L} \bigg]'(q)
			 \Bigg\} +   o(\sigma_{N}^4)   \, .
  \end{eqnarray*}
\end{corollary}
The corresponding result for the expected shortfall is again a direct consequence of Lemma \ref{lem:QuantIntegr}.
\begin{corollary}\label{cor:ES3rdOrdExpan1Dim}
In the one-dimensional case, 
the expansion of  $\ES_\alpha[S(\phi)]$ in the asset volatility
$\sigma \in \{\sigma_{N},\sigma_{lN}\}$
up to forth order as $\sigma \to 0$ is given by
\begin{eqnarray*}
\lefteqn{\ES_\alpha[S(\phi)]  = ES_\alpha[-L]
    +  \frac{\sigma^2}{2\alpha} \cdot (\phi-\qL)^2\cdot f_L(q)}&&\\
    &&+ 
    \left\{
      \begin{array}{lll}
        \begin{array}{l}
        \frac{\sigma^3 \!\cdot\!\mu_3}{6\alpha}\cdot(\phi-q)^3 \cdot f_L'(q)
        +
        \frac{\sigma^4}{24\alpha}\cdot\bigg[  \mu_4\cdot \big[(\phi-id)^4\!\cdot\! f_L'\big]'
			  - 3\cdot\frac{{\big((\phi-id)^2\!\cdot\! f_L\big)'\,}^2}{f_L} \\
				+(2\mu_4- 6) \cdot (\phi-id)^3 \!\cdot\!f_L'
        + \left(\mu_4 +3\right) \cdot (\phi-id)^2 \!\cdot\!f_L	\bigg](q)
        + o(\sigma^4)\\
        \end{array}
        &  &       (\sigma = \sigma_{lN}) \, ,\\ \\
         \begin{array}{l}
         \frac{\sigma^3 \!\cdot\!\mu_3}{6\alpha}\cdot    \big((\phi-id)^3\!\cdot\! f_L\big)'(q)
         +\frac{\sigma^4}{24\alpha}\cdot\bigg[  \mu_4\cdot \big((\phi-id)^4\!\cdot\! f_L\big)''\\
			  - 3\cdot\frac{{\big((\phi-id)^2\!\cdot\! f_L\big)'\,}^2}{f_L} \bigg](q)
			 +   o(\sigma^4)\\
        \end{array}
        &  &  (\sigma = \sigma_N) \, .\\
      \end{array}
    \right.
  \end{eqnarray*}
\end{corollary}

\begin{remark}
  {\rm
In contrast to the value-at-risk case, all correction terms of the 
expansions of $\phi\to \ES_\alpha[S(\phi)]$ up to fourth order have
$\phi^* = q$ as (local) minimum, refer also to Figure \ref{fig:VaRES_vs_phi_highAssetVol}.
This is consistent with Theorem \ref{thm:SpecialPoint_phi_ql} stating that the risk-minimizing asset allocation equals $\qL$ independently of the distribution of $X$ and $L$.
  }
\end{remark}

\section{Numerical Analysis}

\subsection{Univariate Case}

We now compare our perturbation results in the univariate case 
with numerical analysis. To this end we use numerical integration and sample the cumulative distribution function 
of the surplus 
 around the $\alpha$-quantile of the surplus $S(\phi)$ in order to obtain the inverse.
%
%
Figure \ref{fig:VaR_vs_phi} shows the function $\phi\mapsto \rho[S(\phi)]$ for the risk measures $\rho\in\{\VaR_\alpha, \ES_\alpha\}$ with the Solvency II risk tolerance $1\!-\!\alpha = 99.5\%$. The claim size $L$ is normally distributed such that $q = 1$. Log-normal volatility and skew of the asset $X$ are calibrated to typical values of a 30 year discount factor.
It can be seen that the analytical expansion results for the log-normal asset volatility
(Theorem \ref{thm:PerturbVar} and Corollary \ref{cor:ES3rdOrdExpan1Dim})
approximate the numerical behavior quite well. As predicted the risk minimal investment amount in $X$ is around $\phi^* \approx 0.85$ for $\rho = \VaR_\alpha$ and $\phi^* = 1$ for $\rho = \ES_\alpha$, respectively.
%
\begin{figure}[H]
\centering
\includegraphics[width=0.49\textwidth, angle = 0]{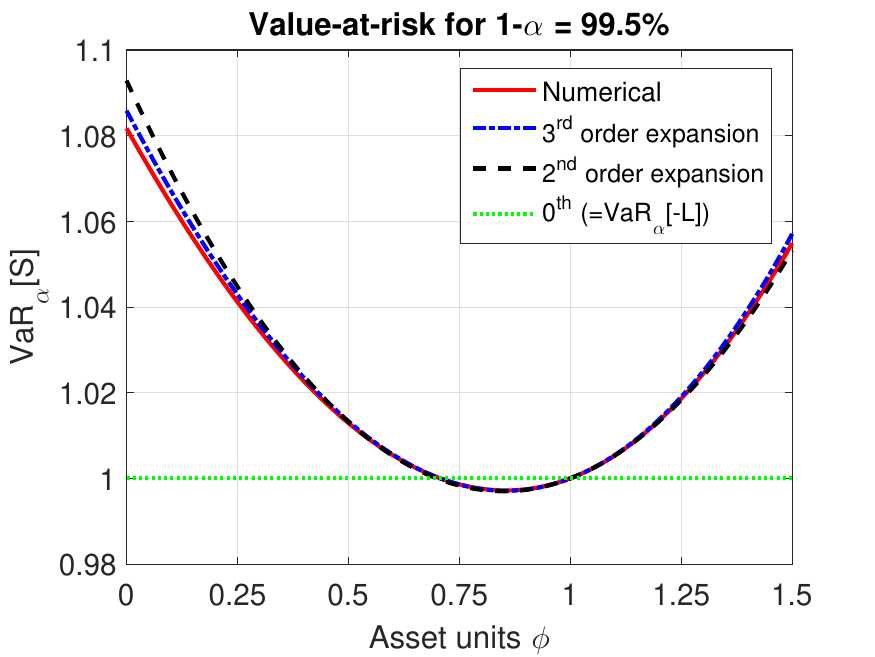}
\includegraphics[width=0.49\textwidth, angle = 0]{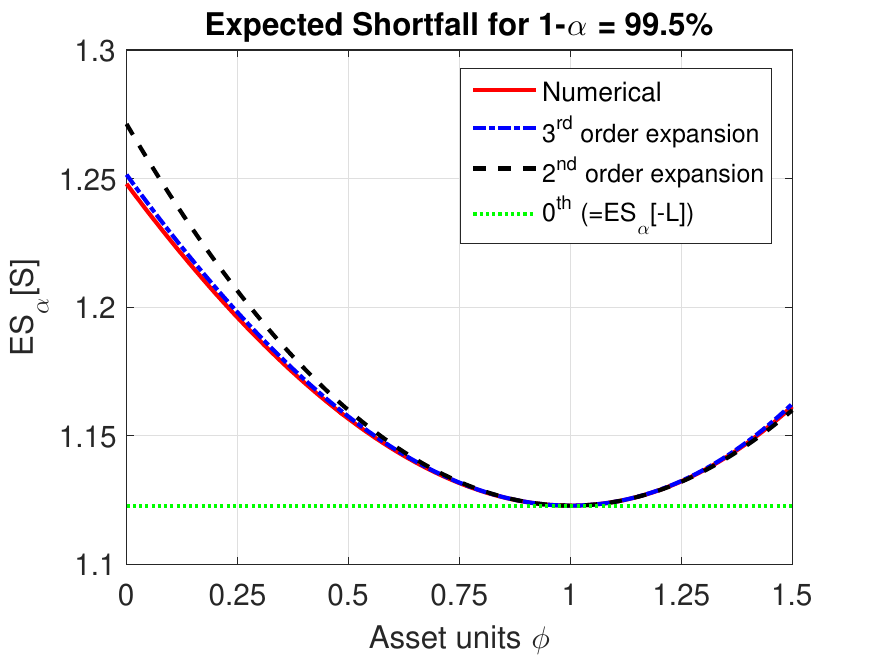}
\caption{\footnotesize Value-at-risk VaR$_\alpha$[S] (left) and expected shortfall ES$_\alpha$[S] (right) as a function of the units $\phi$ of the financial asset $X$. The risk tolerance is set to $1-\alpha = 99.5\%$, the non-hedgeable component $L$ is normally distributed with $\sigma_L=0.388$ such that $q=$VaR$_\alpha$(-L)=1, and $\log(X)$ is log-normally distributed such that $X$ has log-normal volatility $\sigma = 0.2$ and log-normal skew $\mu_3=-0.3$. }
\label{fig:VaR_vs_phi}
\end{figure}
Figure \ref{fig:VaRES_vs_phi_highAssetVol} displays the same situation as Figure \ref{fig:VaR_vs_phi}, but with a much more volatile asset (comparable to an emerging market single stock).
For both risk measures the third and fourth order expansions based on normal asset volatility are less accurate than the
expansions based on log-normal asset volatility.
In the value-at-risk case the second order approximation still fits the overall shape quite well, whereas the third and fourth order expansion are more accurate for investment amounts $\phi$ not too far from $q$;
the optimal investment $\phi^* \approx 0.9$ is higher than in the second order approximation due to the massive
negative asset skew;  in this setting $\phi^*$ is very close to the optimal investment in the third order approximation, whereas the fourth order correction of the optimal investment does not add precision if $\phi$ is away from $q$.
In the expected shortfall case, the third order (log-normal volatility based) approximation produces the best fit for the risk profile, whereas the fourth-order approximation adds only little additional accuracy for $\phi$ not too far from $q$.
These observations are consistent with the fact that the Gram-Charlier series are known to converge slowly, see e.g.~\cite{Milne}.
%
\begin{figure}[H]
\centering
\includegraphics[width=0.45\textwidth, height=7.5cm, angle = 0]{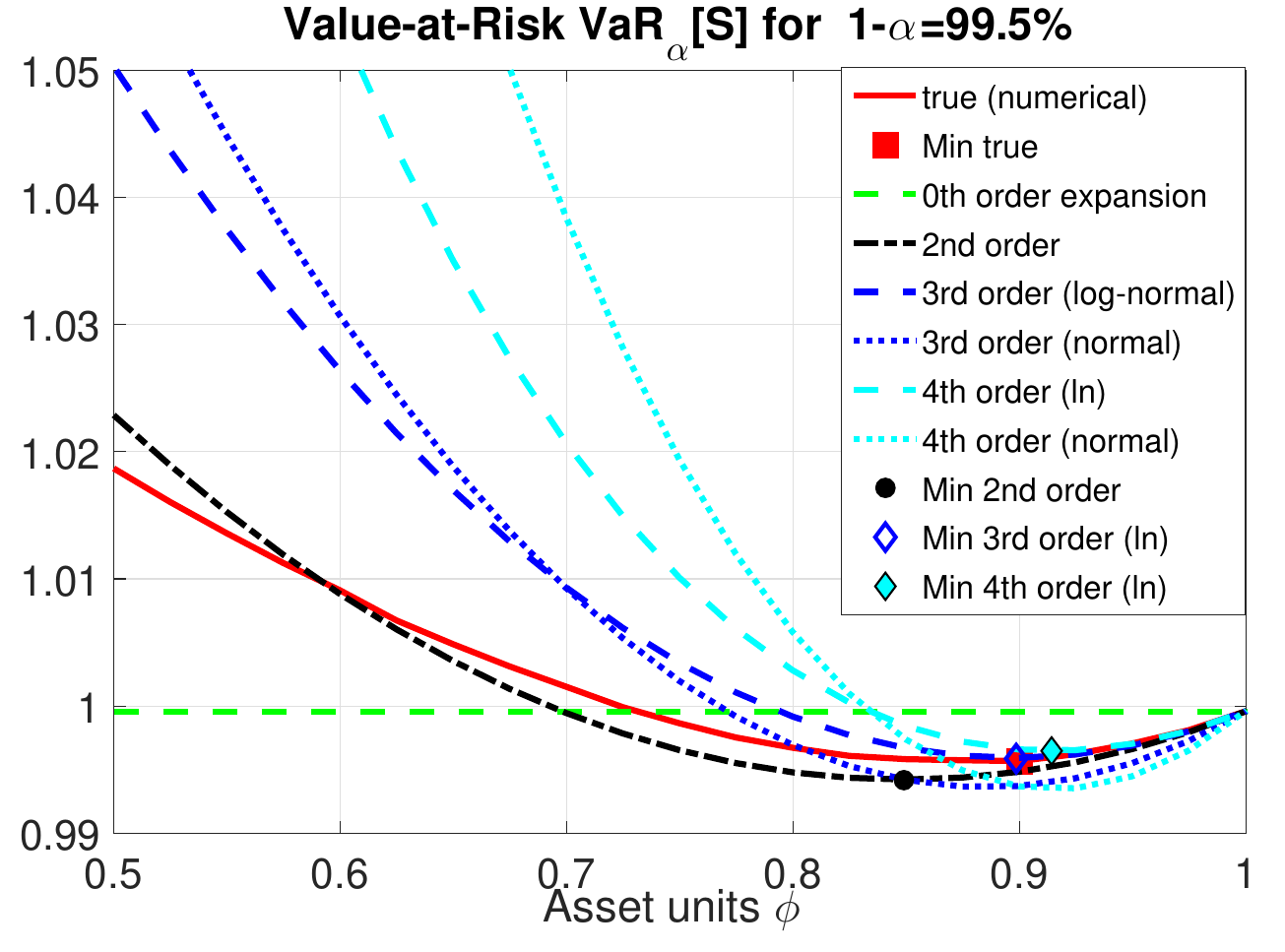}
\hspace{2mm}
\includegraphics[width=0.45\textwidth, height=7.5cm, angle = 0]{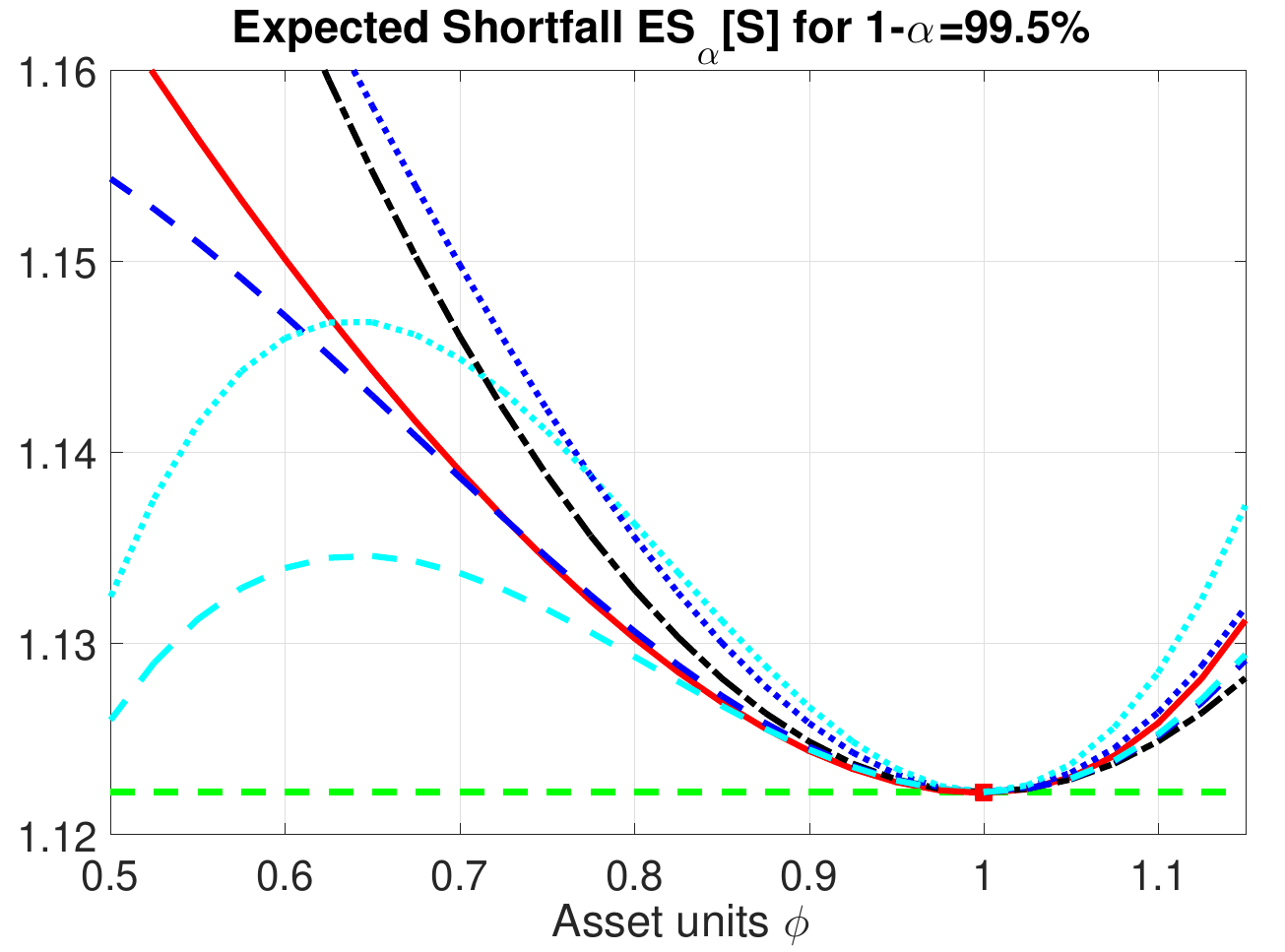}
\caption{\footnotesize Same setting as in Figure \ref{fig:VaR_vs_phi} but much more volatile asset:
log-normal volatility of $\log(X)$ amounts to $\sigma = 0.5$ which implies a log-normal skew $\mu_3=-1.75$. }
\label{fig:VaRES_vs_phi_highAssetVol}
\end{figure}

Next we analyze for the value-at-risk 
the location of the risk minimal investment amount $\phi^*$ in more detail, which depends on the characteristics of the hedgeable risk factor $X$.
Figure \ref{fig:phi*_vs_sigma} shows the dependence of $\phi^*$ on the log-normal volatility $\sigma$ for various log-normal skew values $\mu_3$.
%
\begin{figure}[H]
\centering
\includegraphics[width=0.32\textwidth, height=6.5cm]{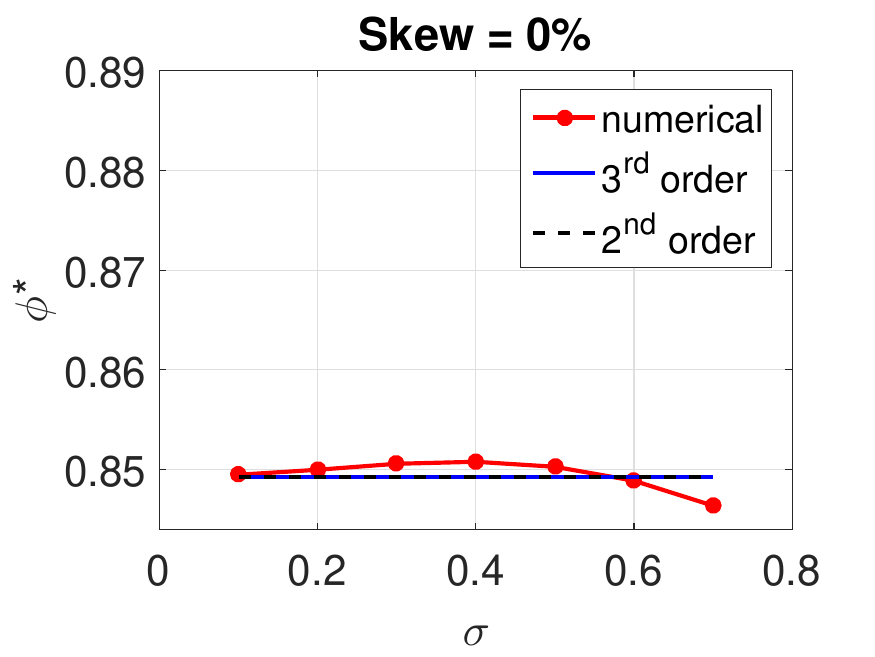}
\includegraphics[width=0.32\textwidth, height=6.5cm]{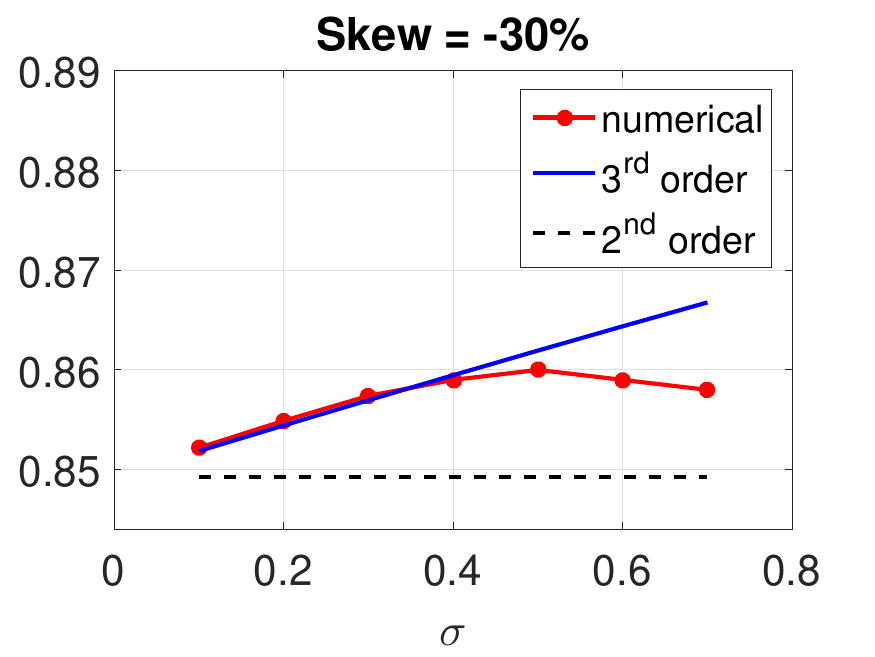}
\includegraphics[width=0.32\textwidth, height=6.5cm]{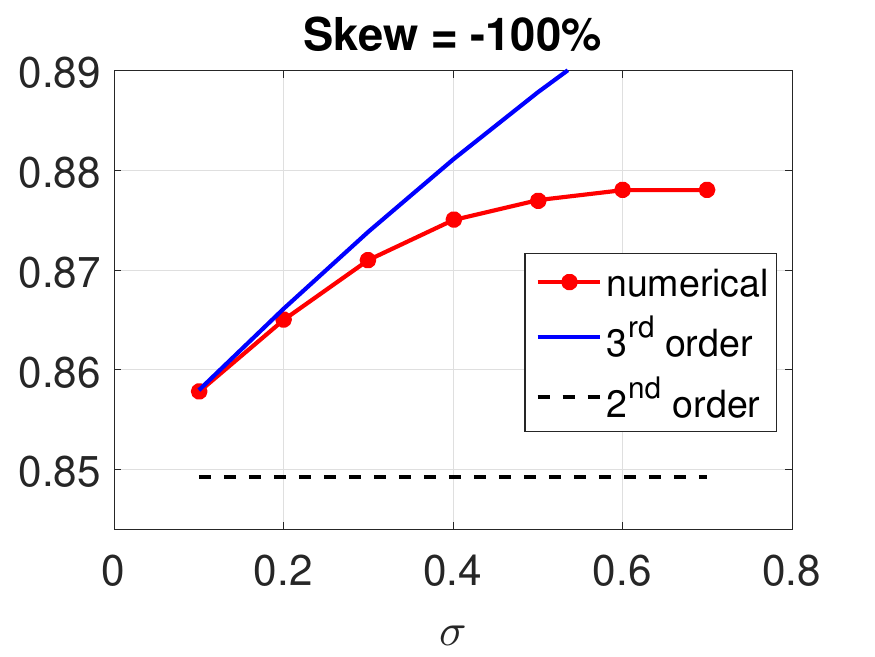}
\caption{\footnotesize Optimal investment amount $\phi^*$ minimizing the value-at-risk VaR$_\alpha$[S($\phi$)] as a function of the log-normal volatility $\sigma$ of the financial asset $X$ for various log-normal skews $\mu_3$. Refer to the description of Figure \ref{fig:VaR_vs_phi} for further calibration details.}
\label{fig:phi*_vs_sigma}
\end{figure}
In case of zero skew the third order expansion term vanishes. Higher order terms lead only to very small corrections to our theoretical prediction of $\phi^* \approx 0.85$. For realistic skew values of around $\mu_3 = -0.3$ the third order expansion is a good approximation up to $\sigma = 0.5$. In case of very high skew $\mu_3 = -1.0$ the approximation is only good up to $\sigma = 0.3$.
To sum up, for realistic parametrizations of the hedgeable risk factor $X$ our perturbation results up to third order reflect the behavior of the risk minimal investment amount $\phi^*$ very well.

\subsection{Bivariate Case}
Next let us consider the case of two financial assets $X_1$ and $X_2$, which are used to hedge two different claim sizes $L_1$ and $L_2$. Based on Monte Carlo simulation we compare the numerical results for the risk minimal investment amounts $\phi^*_1$ and $\phi^*_2$ with the findings of our perturbation approach.

Figure \ref{fig:VaR_vs_phi_2d} shows the numerical results for the value-at-risk $VaR_\alpha[S(\Bphi)]$  as a function of the units $\Bphi=(\phi_1, \phi_2)$ of the financial asset $\BX$. As in the univariate case the risk tolerance is set to $1\!-\!\alpha = 99.5\%$ and the claim size $\BL$ is normally distributed such that $q = 1$. The financial assets $X_1$ and $X_2$ are chosen to be independent and log-normally distributed with log-normal volatility $\sigma = 0.3$. For the symmetric case (a) the analytical expansion results
in second order (Theorem \ref{thm:AllocPhiOptLmultiNormal}) predict risk minimal investment amounts of $\phi^*_1= \phi^*_2=\phi_0^*/2\approx 0.425$. In the asymmetric case (b) we obtain $\phi^*_1 \approx 0.79$ and $ \phi^*_2\approx 0.06$. In both cases the numerical results coincide quite well with the theoretical prediction.
\begin{figure}[H]
\includegraphics[width=0.49\textwidth, angle = 0]{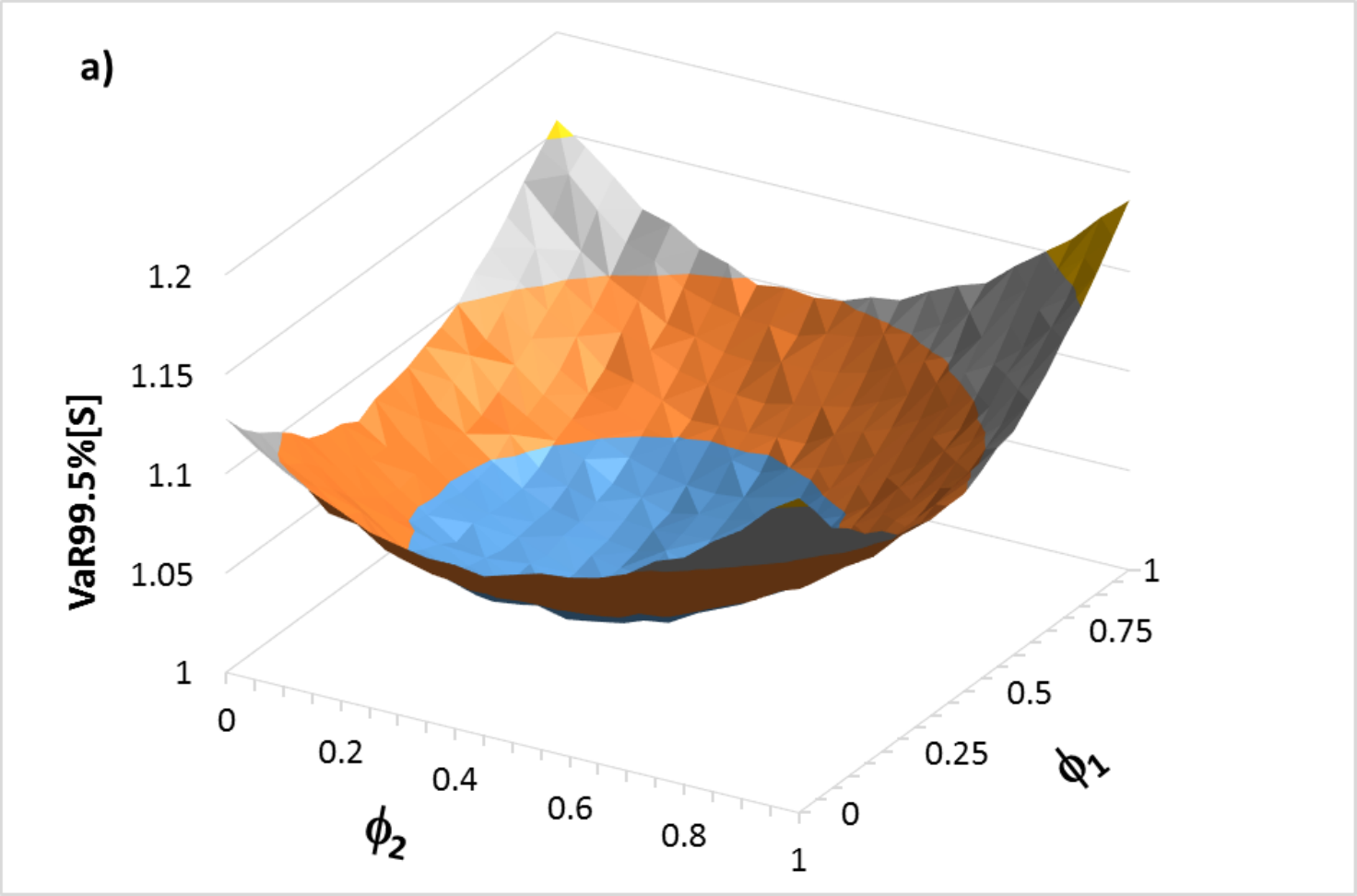}
\includegraphics[width=0.49\textwidth, angle = 0]{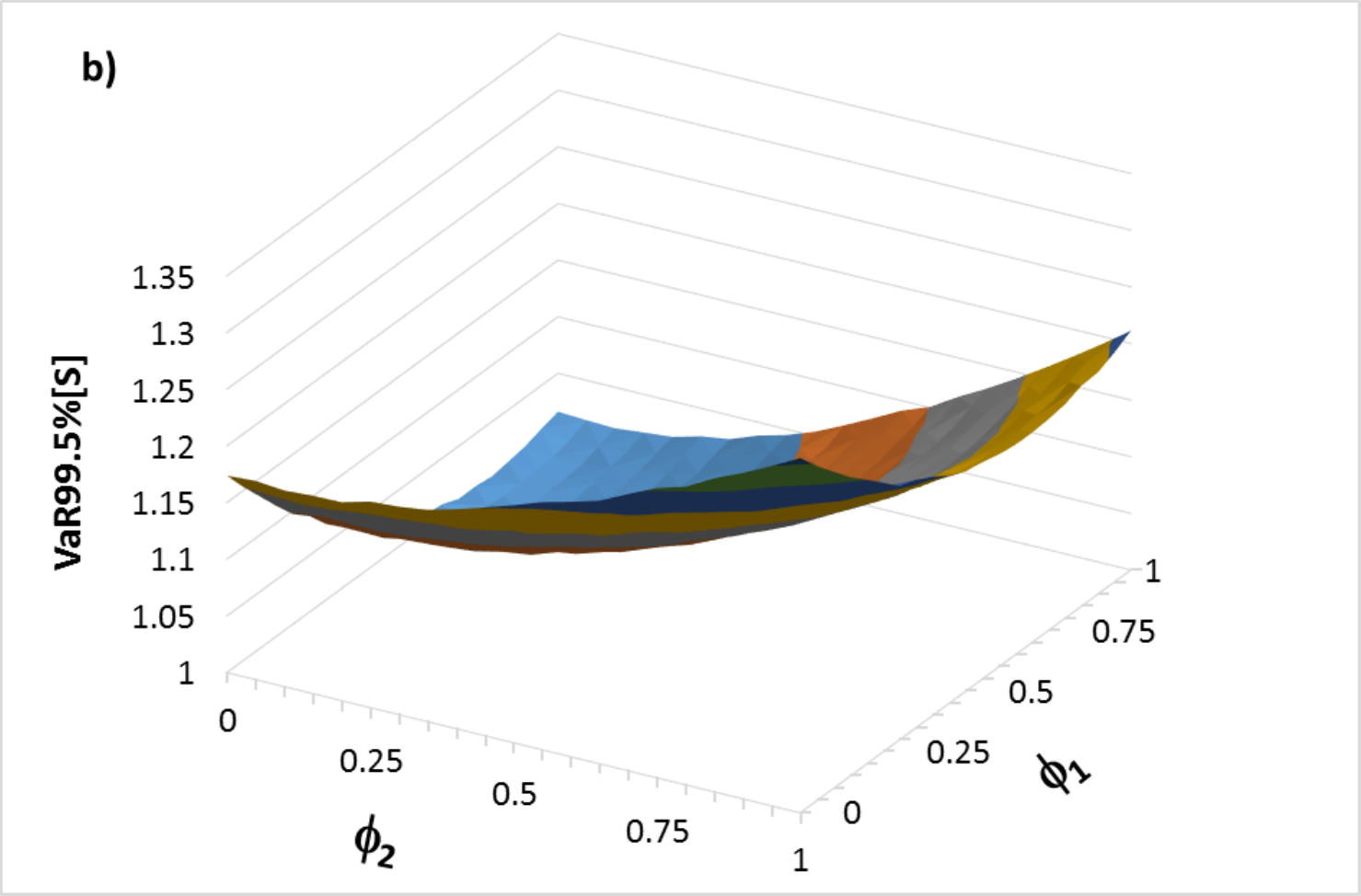}
\caption{\footnotesize Value-at-risk VaR$_\alpha$[S] derived from Monte Carlo simulation as a function of the units $\phi_1$ and $\phi_2$ of the two-dimensional financial asset $\BX$ for risk tolerance $1-\alpha = 99.5\%$. The two financial assets $X_1$ and $X_2$ are independent and log-normally distributed with log-normal volatility $\sigma = 0.3$. The non-hedgeable components $L_1$ and $L_2$ are also independent but normally distributed. In the symmetric case (a) the covariance matrix is set to $\Sigma^L_{11}=\Sigma^L_{22} = 0.0756$ and in the asymmetric case (b) we have $\Sigma^L_{11}=0.141$ and $\Sigma^L_{22} = 0.01$. }
\label{fig:VaR_vs_phi_2d}
\end{figure}

\section{Application to Solvency II Market Risk Measurement}
\label{sec:Application}

In general, there are two ways of how to set up an internal model for calculating the Solvency Capital Requirement (SCR) under Solvency II:
The integrated risk model calculates the surplus (= excess assets over liabilities) distribution of the economic balance sheet, by simulating simultaneously the stochastics of all risk drivers (hedgeable and non-hedgeable). Although it is the more adequate approach, it is rarely used in practice both for operational and steering reasons.
Market standard is a modular approach similar to the one used in the Solvency II standard formula. In the modular risk model the profit and loss distribution for each risk category is computed in a separate module and the different risk modules are subsequently aggregated to the total SCR of the company. For risk categories which affect only one side of the economic balance sheet this approach works fine. The market risk module is more problematic, because risk drivers like foreign exchange rates or interest rates affect both sides of the balance sheet. Therefore so-called replicating portfolios are introduced, which translate the capital market sensitivities of the liability side into a portfolio of financial instruments (e.g. zero coupon bonds). The key question is, how the notional value of the liabilities should be chosen for the replicating portfolio? Market standard is to take the best-estimate value,
which implies that the capital backing the surplus is attributed to the risk-free investment, e.g.~EUR cash.
We will show that this can lead to significant distortions of the measured market risk SCR as compared to an integrated risk model. To avoid this we have introduced at Munich Re the concept of the Economic Neutral Position (ENP)
which is defined as (virtual) asset portfolio, which minimizes the total SCR of the integrated model.
The ENP is the risk-neutral reference point for Solvency II market risk measurement
in Munich Re's certified internal model.\footnote{Except for with-profit life insurance business which exhibits significant interaction between the asset and the liability side of the insurer's balance sheet.}
This means that any mismatch between assets and ENP produces market risk by definition.

}

For liabilities exhibiting the product structure $\sum_i L_i\cdot X_i$ defined in section \ref{sec:Setup}, the ENP corresponds exactly to the solution of the optimization problem addressed in this paper. The ENP consists of  assets $X_i$ (represented by zero coupon bonds of different maturity and currency), which back the claim cash flows of the liability side in a risk minimal way. The investment amounts of the assets in the ENP
equal the best estimate values of $L_i\cdot X_i$
plus a safety margin corresponding to the risk minimal investment amount $\phi^*_i$.
If the $L_i$ are normally distributed then the total safety margin equals 
85\% of the total insurance risk component $\rm SCR_{L_i}$ defined as the risk contribution
for the non-hedgeable claim size $L_i$ fully diversified within all non-hedgeable risks.
This component is allocated to the single assets $X_i$ (e.g.~the different maturities of the zero bonds)
according to the covariance principle (Theorem \ref{thm:AllocPhiOptLmultiNormal}).

Let us now analyze the total SCR of a modular risk model, which uses the ENP as risk-neutral reference portfolio for market risk measurement, and compare it with the outcome of an integrated risk model. We assume that the surplus $S$ is of the form (\ref{eq:DefSphiSimplif}) for the one-dimensional case. Let us consider the Solvency II risk measure $\VaR_\alpha[S]$ with risk tolerance $1\!-\!\alpha = 99.5\%$. The non-hedgeable ${\rm SCR}_L$ of the  insurance liabilities is computed in the insurance 
risk module (e.g. the P/C module). For our simple example ${\rm SCR}_L$ equals our definition of $q$ and can be set to one without loss of generality (${\rm SCR}_L = q = \VaR_\alpha[L] = 1$).
The market risk SCR$_M$ is measured by the value-at-risk of the mismatch portfolio of assets minus ENP, i.e.~$ S_M(\phi) = (\phi - \phi^*) \cdot X  - \phi$, and is a function of the units $\phi$ of the financial asset $X$. For the sake of simplicity the total SCR$_T$ of the surplus is calculated by aggregating SCR$_L$ and SCR$_M$ based on the square root formula, which is also used in the Solvency II standard formula (remember that $L$ and $X$ are assumed to be independent): ${\rm SCR}_T= \sqrt{{\rm SCR}_L^2 + {\rm SCR}_M^2}$. This aggregation method is only valid for a sum of normally distributed stochastic variables. Therefore we assume that both risk drivers $L$ and $X$ follow a normal distribution,
i.e.~we violate here the positivity assumption on $X$ for technical reasons.
Otherwise the aggregation method needs to be adjusted accordingly.

\begin{figure}[H]
\centering
\includegraphics[width=0.6\textwidth]{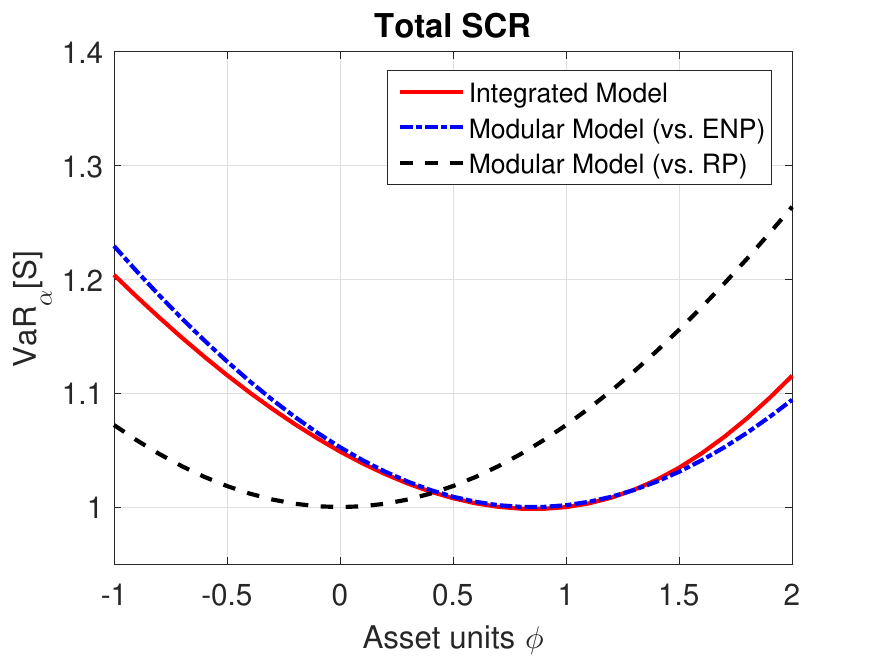}
\caption{\footnotesize Total SCR$_T$  as a function of the units $\phi$ of the financial asset $X$ for an integrated risk model  (red solid) in comparison with a modular risk model, where the market risk is measured either vs. ENP (blue dashed-dotted) or vs. RP (black dashed). $X$ is assumed to be normally distributed with a volatility of 15\%.  }
\label{fig:SCR_total}
\end{figure}

Figure \ref{fig:SCR_total} compares the total SCR$_T$ of the modular risk model with the total SCR$_T$ of the integrated model, which is simply the value-at-risk of $S(\phi)$ at risk tolerance $1\!-\!\alpha = 99.5\%$ with joint stochastics of all risk drivers.
%

The integrated and the ENP-based modular approach
yield in good approximation the same total SCR, as desired. Only if the asset value $\phi$ differs strongly from the risk minimal value $\phi^*$, deviations between the outcomes of the two models can be observed. This is due to the fact, that the square root formula used for aggregation only holds for a sum of normally distributed stochastic variables.
Due to the product structure $L \cdot X$ the total distribution of the surplus is in general not normally distributed  (even though both $L$ and $X$ are normally distributed). This effect can be healed to some extent by refining the aggregation method for the modular model.

For comparison we show in Figure \ref{fig:SCR_total} also the industry standard, which measures market risk versus the replicating portfolio (RP). This corresponds to setting the notional of the liability $L$ equal to its best-estimate value, which is zero in our example. This can lead to substantial deviations from the ``true" SCR as measured by the integrated model.
Especially if the asset amount is below the expected claim size -- a typical case for life insurers whose asset duration
is generally lower than the duration of the liabilities due to the long-term nature of the business -- the modular RP-based approach understates the ``true" risk significantly.





\begin{appendix}

\section{Proofs}

\emph{Justification of the simplifying assumptions (\ref{Ass:WLOG_Assumpt}):}
 $\E[S(\Bphi)] =  \scp{\E[\BX] - \Bx}{\Bphi} + A_0 - \scp{\E[\BX]}{\E[\BL]}$, hence
\begin{eqnarray*}
\lefteqn{S(\Bphi) - \E[S(\Bphi)]
=  \scp{\BX - \E[\BX] }{\Bphi}  - \big(\scp{\BX}{\BL} - \scp{\E[\BX]}{\E[\BL]} \big)}&&\\
&&= \scp{\BX - \E[\BX] }{\Bphi-\E[\BL]}  - \scp{\BX}{\BL-\E[\BL]}
 = \scp{\tilde{\BX} - \Bone}{\tilde{\Bphi}}  - \scp{\tilde{\BX}}{\tilde{\BL}} =: \tilde{S}(\tilde{\Bphi}) \, ,
\end{eqnarray*}
where $\tilde{X}_i := X_i / \E[X_i]$, $\tilde{L}_i :=  \E[X_i] \cdot(L_i - \E[L_i])$, and
$\tilde{\phi}_i :=  \E[X_i] \cdot(\phi_i - \E[L_i])$.
If $\E[\BX] = \Bx$, the cash invariance property of the risk measure yields
$\rho[S(\phi)] = \rho[\tilde{S}(\tilde{\Bphi})] + A_0 - \scp{\E[\BX]}{\E[\BL]}$.
If $\E[\BX] \neq \Bx$, the additional linear term $\scp{\E[\BX] - \Bx}{\Bphi}$ appears.
\\

\emph{Proof of Lemma \ref{lem:PrelimResultsFromAssump}:}
Set $G(\Bphi,z):= \Pa\left(S(\Bphi)\leq z\right)
= \E_\BX \left[ \int_{\{\Bl\in\R^n: \scp{\BX}{\Bl}  \geq \scp{\BX - \Bone}{\Bphi} -  z\}} f_\BL(\Bl) d\Bl  \right]$.
Changing to the rotated variable $\Blambda = (\lambda_1,\bar{\Blambda})'$ defined by $\Bl = \BD\Blambda$ as in
Theorem \ref{thm:RepresK_Terms}, which implies
$\scp{\BX }{\BD\Blambda } = \frac{\lambda_1}{\sqrt{n}}\scp{\BX}{\Bone}  + \scp{\BX}{\BoneT \bar{\Blambda}}$, we obtain
$G(\Bphi,z) = \E_\BX \left[ \int_{\R^{n-1}} d\bar{\Blambda} \int_{\frac{\sqrt{n}}{\scp{\BX}{\Bone}} v}^\infty d\lambda_1 \, g(\lambda_1,\bar{\Blambda}) \right]$, where
$v = v(\BX, \bar{\Blambda},z,\Bphi) := \scp{\BX-1}{\Bphi} - z - \scp{\BX}{\BoneT\bar{\Blambda}}$ and $g(\Blambda) := f_\BL(\BD \Blambda)$ is the rotated density.
The differentials $D_y$ of $G$ with $y \in \{z,\phi_1, \dots, \phi_n\}$ read
$D_y G(\Bphi,z) = -\E_\BX \left[ \int_{\R^{n-1}} g(\frac{\sqrt{n}}{\scp{\BX}{\Bone}} v,\bar{\Blambda}) \cdot \frac{\sqrt{n}}{\scp{\BX}{\Bone}} D_y v \, d\bar{\Blambda} \right]$, where $ D_y v = -1$ if $y = z$ and $= X_i -1$ if $y = \phi_i$. Differentiation and integration can be interchanged by dominated convergence as the (rotated) density $g$ of $\BL$ is bounded and $1/\scp{\BX}{\Bone}$ is integrable by assumption.
Note that the partial derivatives of $G$ are continuous, which implies that the total differential of $G$ exists.
In particular, $z\mapsto G(\Bphi,z)$ is continuous and is an increasing function with $G(\Bphi,\R) = [0,1]$.
Hence for every $\Bphi \in \R_+^n$ and $\alpha\in[0,1]$ there exists a unique $z_{\phi,\alpha}\in \R$ such that
$\Pa(S(\Bphi)\leq z_{\phi,\alpha}) = G(\Bphi,z_{\phi,\alpha}) =\alpha$, which proves (a).
The latter also implies that $S(\phi)$ has  no atoms, and hence upper and lower quantile of $S(\Bphi)$ coincide; the representation for the expected shortfalls follows from Corollary 4.49 of \cite{FoellmerSchied}, hence (b) is proved.\\
Ad (c): since $G$ is continuously differentiable and $D_z G >0$ by the strict positivity of the density of $\BL$, the implicit function theorem implies that $\phi\mapsto z_{\phi,\alpha}$ is differentiable.
For the expected shortfall the differentiability with respect to $\phi_i$ follows from
the representation 
$\ES_\alpha[S(\Bphi)]=\alpha^{-1}\cdot\int_0^\alpha \VaR_\beta[S(\phi)] \, d\beta$, since the differential $\partial_{\phi_i}$ and the integral $\int_0^\alpha$ can be interchanged. This proofs (c)\\
Ad (d): for $\Bphi_1, \Bphi_2 \in \R_+^n$ and $\lambda \in [0,1]$,
\begin{eqnarray*}
\lefteqn{ S\big(\lambda\cdot\Bphi_1 +(1-\lambda)\cdot\Bphi_2\big)
= \scp{\BX-\Bone}{\lambda\cdot\Bphi_1 +(1-\lambda)\cdot\Bphi_2} - \scp{\BX}{\BL}}&&\\
&&=\lambda\cdot [\scp{\BX-\Bone}{\Bphi_1} - \scp{\BX}{\BL}]  +(1-\lambda)\cdot[\scp{\BX-\Bone}{\Bphi_2} - \scp{\BX}{\BL}]
=\lambda\cdot S(\Bphi_1) +(1-\lambda)\cdot S(\Bphi_2) \, .
\end{eqnarray*}
Hence the assertion follows from the convexity of the expected shortfall.
\\

\emph{Proof of part (b) of Theorem \ref{thm:SpecialPoint_phi_ql}:}
In the one-dimensional case, the cumulative distribution of the surplus can be written
\begin{eqnarray}
  F_{S(\phi)}(z)  &=& \Pa\left(\phi\cdot(X-1)-X\cdot L \leq z\right)
   = \E_X \left[ \Pa\left(L \geq \phi-(z+\phi)/X \, \big| \, X \right) \right] \nonumber\\
   &=& \E_X \left[ \bar{F}_L \Big(w(\phi,z,X) \Big) \right] \, , \qquad    w(\phi,z,X) := \phi-(z+\phi)/X \, ,\label{eq:ReformulFSphi}
\end{eqnarray}
where the last two equations follow from the strict positivity of $X$ and its independence from  $L$.
Since the quantile $z_\phi$ is implicitly defined as the $z$ solving
$\alpha = F_{S(\phi)}(z) =
\E_X \left[ \bar{F}_L \Big(w(\phi,z,X) \Big) \right]$, we
can determine $\partial_\phi z_\phi$ at $\phi = q$ from the implicit function theorem (whose conditions are satisfied as shown in proof of Lemma \ref{lem:PrelimResultsFromAssump}).
%
We denote by $D_\phi = \partial_\phi + (\partial_\phi z_\phi)\cdot \partial_z$ the total differential with respect to $\phi$.
Applying $D_\phi$ on the defining equation of $z_\phi$
yields
\begin{equation}
  0 = D_\phi\, \E_X\big[\bar{F}_L(w(\phi,z_\phi,X)) ]
  = -\E_X\big[f_L(w(\phi,z_\phi,X)) \cdot
  [\partial_\phi + \partial_\phi z_\phi\cdot \partial_z]
  w (\phi,z_\phi,X)\big]\label{eq:EfLw_Dphi_w_IsZero}
\end{equation}
Since $\partial_\phi w = 1-1/X$ and $\partial_z w = -1/X$
we deduce
\begin{equation*}
  \partial_\phi z_\phi = \frac{\E_X\big[f_L( w) \cdot(1-1/X)\big]}{\E_X\big[f_L( w) \cdot(1/X)\big]}
  = \frac{\E_X\big[f_L( w)\big]}{\E_X\big[f_L( w) \cdot(1/X)\big]} -1\, ,
  \end{equation*}
provided the denominator is not zero.
Since $z_q = -q$, the term 
$w(q,z_q,X) = q-(q+z_q)/X = q$ becomes constant. Hence also $f( w)$ becomes constant and the expression for $\partial_\phi z_\phi$ above collapses to
\begin{equation}
  (\partial_\phi z_\phi)_{|_{\phi=\qL}} = \E[X^{-1}]^{-1} -1 \leq 0 \, ,\label{eq:partialZphiIsqL}
\end{equation}
with $<$ if $X$ is non constant. The latter inequality follows from the strict convexity of the inverse function and Jensen's inequality, which implies $\E[X^{-1}] > \E[X]^{-1} = 1$ for non-constant $X$.
Multiplying (\ref{eq:partialZphiIsqL}) with $-1$ yields the assertion of the theorem for the value-at-risk.

For the expected shortfall, we can show that at $\phi = \qL$ the derivative with respect to $\phi$ vanishes: from the second equation in (\ref{eq:ReformulFSphi}) we find that $\{S(\phi) \leq z_\phi\} = \{L \geq w(\phi,z_\phi,X)\}$.
Similar to (\ref{eq:reformESSphi}) we calculate
\begin{eqnarray*}
  \E[S(\phi)\cdot \one_{S(\phi)\leq z_\phi}]
	&=& \E_X\Big[ \big(\phi\cdot(X-1) -X\cdot L\big) \cdot \one_{L\geq w(\phi,z_\phi,X)}\Big] \\
   &=& \phi\cdot \E_X\Big[ (X-1) \cdot \bar{F}_L\big( w(\phi,z_\phi,X)\big)\Big] - \E_X\Big[X \cdot
	\int_{w(\phi,z_\phi,X)}^\infty l\cdot f_L(l) \, dl\Big]\, .
\end{eqnarray*}
Differentiation with respect to $\phi$ yields
\begin{eqnarray*}
 \partial_\phi 
 \E[S(\phi)\cdot \one_{S(\phi)\leq z_\phi}]
   &=& \E_X[ (X-1) \cdot \bar{F}_L(w)] - \phi\cdot \E_X[ (X-1) \cdot f_L(w)\cdot D_\phi w] \\
   &+& \E_X[X \cdot w\cdot f_L(w)\cdot D_\phi w] \, .
\end{eqnarray*}
Recall that at $\phi = \qL$, the term $w(\qL,z_\qL,X) = \qL$ becomes constant. Hence the above expression simplifies
\begin{eqnarray*}
\partial_\phi \E[S(\phi)\cdot \one_{S(\phi)\leq z_\phi}]_{|_{\phi=\qL}}
   &=& \bar{F}_L(\qL)\cdot\E_X[ X-1] + \qL \cdot f_L(\qL)\cdot \E_X\Big[ \big(-(X-1)+X\big) \cdot D_\phi w\Big] \\
   &=& \qL \cdot f_L(\qL)\cdot \E_X[ (D_\phi w)(\qL,z_\qL,X)] = 0 \, ,
\end{eqnarray*}
where the last equality follows from the unit-mean property of $X$ and from (\ref{eq:EfLw_Dphi_w_IsZero}) evaluated at $\phi=\qL$ together with the fact that $f_L(w)$ becomes a positive constant.
This proves the assertion of the theorem for the expected shortfall.
\\


{\em Proof of Proposition \ref{prop:ExpansSum2depRVs}:}
The characteristic function of $Y_0 + Y_1$ 
can be written as
$\phi_{Y_0 + Y_1}(t) := \E[e^{it(Y_0 + Y_1)}]
= \E_{Y_1}\big[e^{itY_1} \cdot\phi_{Y_0 | Y_1}(t) \big]$,
where $\phi_{Y_0 | Y_1}(t) := \E[e^{itY_0} | Y_1]$ denotes the conditional characteristic function of $Y_0$ conditioned on
$Y_1$.

We show that $\phi_{Y_0 + Y_1}$ and $\phi_{Y_0 | Y_1}$ are integrable: by assumption the differential of any order of the density $f_{Y_0+Y_1}$ exists and is integrable.
Since $f_{Y_0+Y_1}$ is continuous and hence locally bounded, it is also $L^2$-integrable. We deduce from Parceval's theorem and the differentiation rules for the Fourier transformation that
$\int_\R |D^k f_{Y_0+Y_1}|^2 \, dx = \frac{1}{\sqrt{2\pi}}
\int_\R |t^k \cdot\phi_{Y_0+Y_1}(t)|^2 \, dt$ for every $k \in\N_0$.
As any characteristic function is bounded, $\phi_{Y_0+Y_1}$
is integrable since the tails are integrable by Cauchy-Schwartz:
$\int_{T_0}^\infty |\phi_{Y_0+Y_1}| \, dt
\leq (\int_{T_0}^\infty t^{-2}\, dt) \cdot (\int_{T_0}^\infty t^2|\phi_{Y_0+Y_1}| \, dt < \infty$, and analogously for the negative tail.
Since $F_{Y_0+Y_1}(z) = \E_{Y_1}[F_{Y_0| Y_1} (z-Y_1 )]$, the
differentiability- and integrability-assumptions for $F_{Y_0+Y_1}$ also hold for the conditional cumulative distribution $F_{Y_0| Y_1}$.
Repeating the above arguments, we deduce that
$\phi_{Y_0 | Y_1}$ is also integrable.
%

By the inversion formula, the cumulative distribution of $Y_0+Y_1$ can be recovered for
$z_0<z$
\begin{eqnarray*}
F_{Y_0 + Y_1}(z) - F_{Y_0 + Y_1}(z_0)
&=& (2\pi)^{-1} \int_\R \frac{ e^{-itz_0}-e^{-itz}}{it}\cdot \phi_{Y_0 + Y_1}(t) \, dt\\
&=& (2\pi)^{-1} \int_\R \frac{ e^{-itz_0}-e^{-itz}}{it} \cdot \E_{Y_1}\big[e^{itY_1} \cdot\phi_{Y_0 | Y_1}(t) \big] \, dt\\
&=& (2\pi)^{-1} \E_{Y_1}\left[ \, \int_\R \sum_{r = 0}^\infty \frac{(itY_1)^r}{r!}
		\cdot \frac{ e^{-itz_0}-e^{-itz}}{it} \cdot\phi_{Y_0 | Y_1}(t) \, dt\right]\\
&=& (2\pi)^{-1} \sum_{r = 0}^\infty \frac{(-1)^r}{r!} \cdot \E_{Y_1}\left[ Y_1^r \int_\R (-it)^r
		\cdot \frac{ e^{-itz_0}-e^{-itz}}{it} \cdot\phi_{Y_0 | Y_1}(t) \, dt\right] \\
&=& \sum_{r = 0}^\infty \frac{(-1)^r}{r!} \cdot \E_{Y_1}\left[ Y_1^r \cdot \big( D_z^r F_{Y_0|Y_1}(z) - D_z^r F_{Y_0|Y_1}(z_0) \big) \right]\, ,
\end{eqnarray*}
where the third equation follows from Fubini's theorem (since $(t,y_1)\mapsto \phi_{Y_0 | y_1}(t)$ is integrable on the product measure) and from expanding $e^{itY_1}$; the fourth equation follows from the fact that the convergence of the exponential series is uniform on $\{w \in \C : \Re w \leq 1 \}$ and the last equation follows from the differentiation rules for Fourier transforms.
Letting $z_0$ tend to $-\infty$ we obtain
$$F_{Y_0 + Y_1}(z)
= \sum_{r = 0}^\infty \frac{(-1)^r}{r!} \cdot D_z^r \E_{Y_1}\left[ Y_1^r \cdot F_{Y_0|Y_1}(z) \right]
= \sum_{r = 0}^\infty \frac{1}{r!} \cdot (-D_z)^r \, \E_{Y_1}\Big[ Y_1^r \cdot \E[\one_{Y_0\leq z}|Y_1]\Big] \, ,$$
which proves the assertion.\\

{\em Proof of Theorem \ref{thm:AllocPhiOptLmultiNormal}:}
We start with some preparations. Since $\BL \sim \Ncal(\Bzero, \BSigmaL)$, also
$(\BL, \scp{\Bone}{\BL})$ is distributed according to a centered $(n+1)$-dimensional normal distribution with covariance matrix
$\Gamma =
\left(
\begin{array}{c | c}
\BGamma_{11}  & \BGamma_{12}\\
\hline
\BGamma_{12}'  & \Gamma_{22}
\end{array}
\right) , $
with
$\BGamma_{11} = \Sigma^\BL$,
$\BGamma_{12} = \Sigma^\BL \cdot \Bone$,
and $\Gamma_{22} = \scp{\Bone}{\BSigmaL \!\cdot\!\Bone}$.
From the theory of conditional normal distributions we derive that
$\BL$ conditioned on the event $\{\scp{\Bone}{\BL} = x \}$
follows a $n$-dimensional normal distribution 
$$\BY := \BL\, | _{\, \{\scp{\Bone}{\BL} = x \}}  \sim
\Ncal\left( \frac{x \cdot \BGamma_{12}}{\Gamma_{22}} \, , \,
\BGamma_{11} - \frac{\BGamma_{12} \cdot \BGamma_{12}'}{\Gamma_{22}}  \right)
=\Ncal\left(
\frac{x \cdot \BSigmaL \!\cdot\!\Bone }{\scp{\Bone}{\BSigmaL \!\cdot\!\Bone}} \, , \,
 \Sigma^\BL - \frac{(\Sigma^\BL \!\cdot\!\Bone) \cdot (\Sigma^\BL \!\cdot\!\Bone)'}{\scp{\Bone}{\BSigmaL \!\cdot\!\Bone}} \right) \, .$$
Hence
\begin{eqnarray*}
\E[\BL  \, | \,  \scp{\Bone}{\BL} ]
&=& \E[\BY]= {\scp{\Bone}{\BSigmaL \!\cdot\!\Bone}}^{-1}
\cdot \scp{\Bone}{\BL} \cdot \BSigmaL \!\cdot\!\Bone \, ,\\
\E[L_i\!\cdot\! L_j  \, | \,  \scp{\Bone}{\BL} ] &=&
 \E[Y_i\!\cdot\! Y_j]
= \E\Big[(\BY - \E[\BY])_i \cdot(\BY - \E[\BY])_j\Big] + \E[Y_i]\cdot\E[Y_j]
\\
&=& \BSigma^\BL_{ij} - 
{\scp{\Bone}{\BSigmaL \!\cdot\!\Bone}}^{-1} \cdot (\BSigma^\BL \!\cdot\!\Bone)_i \cdot (\BSigma^\BL \!\cdot\!\Bone)_j
+ 
{\scp{\Bone}{\BL}^2} \cdot {\scp{\Bone}{\BSigmaL \!\cdot\!\Bone}^{-2}}  \cdot  (\BSigma^\BL \!\cdot\!\Bone)_i \cdot (\BSigma^\BL \!\cdot\!\Bone)_j \, .
\end{eqnarray*}
Denoting the K-terms of the associated single-asset case by $\Ksa_i(q) :=
\E[\scp{\Bone}{\BL}^i \cdot \one_{\scp{\Bone}{\BL}>q}]$ we deduce
\begin{eqnarray}
\BK(q) &=& \E[\BL \cdot \one_{\scp{\Bone}{\BL}>q}]
=\E\big[\E[\BL |\scp{\Bone}{\BL}] \cdot \one_{\scp{\Bone}{\BL}>q}\big]
= \frac{\Ksa_1(q)}{\scp{\Bone}{\BSigmaL \Bone}} \cdot \BSigmaL \Bone \, ,\label{eq:PrfThm13K}\\
\nonumber\\
\BK_\BSigma[\BL](q) &=& \E[\scp{\BL}{\BSigma\cdot\BL} \cdot \one_{\scp{\Bone}{\BL}>q}]
=\E\big[\E[\scp{\BL}{\BSigma\cdot\BL} |\scp{\Bone}{\BL}] \cdot \one_{\scp{\Bone}{\BL}>q}\big]\nonumber\\
&=& \tr(\BSigma \cdot \BSigmaL) \cdot\bar{F}_{\scp{\Bone}{\BL}}(q)
-   \frac{\scp{\Bone}{\BSigmaL \cdot\BSigma\!\cdot\! \BSigmaL \!\cdot\! \Bone}}{{\scp{\Bone}{\BSigmaL \!\cdot\!\Bone}}}
 \cdot \Big(\bar{F}_{\scp{\Bone}{\BL}}(q)
-   \frac{\Ksa_2(q)}{\scp{\Bone}{\BSigmaL \!\cdot\!\Bone}} \Big) \, .
\label{eq:PrfThm13KL}
\end{eqnarray}

{Ad a):} Value-at-risk case:
combining Theorem \ref{thm:PerturbVarMultiDim}.(b) with equation
\eqref{eq:PrfThm13K} gives
$$\Bphi^* = - \frac{\BK''(q)}{ f_{\scp{\Bone}{\BL}}'(q)}
= - \frac{\Ksa_1''(q)}{ f_{\scp{\Bone}{\BL}}'(q)}\!\cdot\!
\frac{\BSigmaL \!\cdot\!\Bone}{\scp{\Bone}{\BSigmaL\!\cdot\! \Bone}} =
\phi_0^* \cdot\frac{\BSigmaL \!\cdot\!\Bone}{\scp{\Bone}{\BSigmaL \!\cdot\!\Bone}} \, , $$
which proves the assertion. The expected shortfall case follows similarly.

{Ad b):}
Value-at-risk case: according to Theorem \ref{thm:PerturbVarMultiDim}.(b) using
\eqref{eq:PrfThm13K} and \eqref{eq:PrfThm13KL}
\begin{eqnarray*}
\VaR_\alpha[S(\Bphi^*)] &=& \qL + \frac{1}{2 f_{\scp{\Bone}{\BL}}(q)} \cdot
	\Big\{	f_{\scp{\Bone}{\BL}}' (q)^{-1} \!\cdot\! \scp{\BK''(\qL) }{\BSigma \cdot \BK''(\qL) } + K_\BSigma[\BL]'' (q)	\Big\}\\
&=& \qL + \frac{1}{2 f_{\scp{\Bone}{\BL}}(q)} \cdot
	\Bigg\{
 \frac{\scp{\Bone}{\BSigmaL \!\cdot\!\BSigma\!\cdot\! \BSigmaL \!\cdot\! \Bone}\cdot\Ksa_1''(q)^2}{{\scp{\Bone}{\BSigmaL \!\cdot\!\Bone}}^2 \cdot f_{\scp{\Bone}{\BL}}'(q)} -  \tr(\BSigma \cdot \BSigmaL) \cdot f_{\scp{\Bone}{\BL}}'(q)\\
&&+  \frac{\scp{\Bone}{\BSigmaL \!\cdot\!\BSigma\!\cdot\! \BSigmaL \!\cdot\! \Bone}}{{\scp{\Bone}{\BSigmaL\!\cdot\! \Bone}}} \cdot \Big(
f_{\scp{\Bone}{\BL}}'(q) + \frac{\Ksa_2''(q)}{\scp{\Bone}{\BSigmaL \!\cdot\!\Bone}}
\Big)\Bigg\}\\
&=& \qL + \frac{f_{\scp{\Bone}{\BL}}'(q)}{2 f_{\scp{\Bone}{\BL}}(q)} \cdot
\Bigg\{
\frac{\scp{\Bone}{\BSigmaL \!\cdot\!\BSigma\!\cdot\! \BSigmaL \!\cdot\!\Bone}}{{\scp{\Bone}{\BSigmaL \!\cdot\!\Bone}}}
-  \tr(\BSigma \cdot \BSigmaL) \\
&&+  \frac{\scp{\Bone}{\BSigmaL \!\cdot\!\BSigma\!\cdot\! \BSigmaL \!\cdot\!\Bone}}{{\scp{\Bone}{\BSigmaL \!\cdot\!\Bone}}^2\cdot f_{\scp{\Bone}{\BL}}'(q)} \cdot \Big(
\frac{\Ksa_1''(q)^2}{f_{\scp{\Bone}{\BL}}'(q)} + \Ksa_2''(q)
\Big)\Bigg\}
 \,  ,
\end{eqnarray*}
which proves the assertions using the fact that
$\frac{\Ksa_1''(q)^2}{f_{\scp{\Bone}{\BL}}'(q)} + \Ksa_2''(q)
=\frac{f_{\scp{\Bone}{\BL}}(q)^2}{f_{\scp{\Bone}{\BL}}'(q)}$, refer also to
\eqref{eq:help_jKjmin1Prim}.

Expected shortfall case:
according to Corollary \ref{corr:PerturbESMultiDim}.(b) using
\eqref{eq:PrfThm13K} and \eqref{eq:PrfThm13KL}
\begin{eqnarray*}
\ES_\alpha[S(\Bphi^*)] &=& \ES_\alpha[-\scp{\Bone}{\BL}]
  - \mbox{$\frac{1}{2\alpha}$} \Big\{	f_{\scp{\Bone}{\BL}} (q)^{-1} \!\cdot\! \scp{\BK'(\qL) }{\BSigma \cdot \BK'(\qL) } + K_\BSigma[\BL]' (q)	\Big\}\\
&=& \ES_\alpha[-\scp{\Bone}{\BL}]
  - \mbox{$\frac{1}{2\alpha}$} \Big\{
 \frac{\scp{\Bone}{\BSigmaL \cdot\BSigma\cdot \BSigmaL , \, \Bone}\cdot\Ksa_1'(q)^2}{{\scp{\Bone}{\BSigmaL \Bone}}^2 \cdot f_{\scp{\Bone}{\BL}}(q)} -  \tr(\BSigma \cdot \BSigmaL) \cdot f_{\scp{\Bone}{\BL}}(q)\\
&&+  \frac{\scp{\Bone}{\BSigmaL \cdot\BSigma\cdot \BSigmaL , \, \Bone}}{{\scp{\Bone}{\BSigmaL \Bone}}}
 \cdot \Big( f_{\scp{\Bone}{\BL}}(q)
+   \frac{\Ksa_2'(q)}{\scp{\Bone}{\BSigmaL \Bone}} \Big)\Big\} \, ,
\end{eqnarray*}
which proves the assertions recalling that
$-\Ksa_2'(q)= \Ksa_1'(q)^2/f_{\scp{\Bone}{\BL}}$.\\

{\em Proof of Equation \eqref{eq:MomentsXlognormExpans}:}
The non-centered i-th moment of $X_{\sigma_{lN}}$ is given by
${m}_i (\sigma_{lN}) := \E[X_{\sigma_{lN}}^i] = M(i\sigma_{lN})/M(\sigma_{lN})^i$.
The moment generating function of $Y$ has the expansion
$M(\sigma) = 1+\mu_2\sigma^2/2 + \mu_3\sigma^3/6 + \mu_3\sigma^4/24 +o(\sigma^4)$ as $\sigma\to 0$, where $\mu_i$ are the moments of $Y$.
Further $(1+x)^{-i} = 1 - ix + i(i+1) x^2/2 + o(x^2)$ as $x\to 0$.
Hence we can write having in mind that $\mu_2=1$ by construction of $Y$
\begin{eqnarray*}
m_i(\sigma_{lN}) &=& \big[1+(i\sigma_{lN})^2/2 + \mu_3(i\sigma_{lN})^3/6 + \mu_4(i\sigma_{lN})^4/24\big] \cdot \\
&&\cdot \big[1- i(\sigma_{lN}^2/2 + \mu_3\sigma_{lN}^3/6+ \mu_4\sigma_{lN}^4/24) + i(i+1)\sigma_{lN}^4/8\big] + o(\sigma_{lN}^4)\\
&=& 1 + i (i-1) \sigma_{lN}^2 / 2 + \mu_3 i(i^2-1)\sigma_{lN}^3/6
+ i\big( \mu_4 (i^3-1) - 6i^2+ 3i+3\big) \sigma_{lN}^4/24+ o(\sigma_{lN}^4) \, .
\end{eqnarray*}
The assertion of \eqref{eq:MomentsXlognormExpans} follows by applying the rule to derive the centered moments $\bar{m}_i$ from the non-centered $m_i$ via
$\bar{m}_i = \sum_{k=0}^i {i \choose k} (-1)^{k- i} m_k $.
\\
%

{\em Proof of Theorem \ref{thm:PerturbVar}:}
Expanding the relation \eqref{eq:n1CumDistSurplExp} up to fourth order in $\sigma \in \{\sigma_{N},\sigma_{lN}\}$
in a similar way as for the derivation of \eqref{eq:SecondOrdExpansStart} having relation \eqref{eq:sigExpansFtail1L_} in mind and omitting the zero and first order terms (which add up to zero by construction) yields
\begin{eqnarray*}
0&=&-f_{L}(-z_0) \cdot(-z_2 -z_3 - z_4)
- \rfrac{1}{2}\cdot f_{L}'(-z_0) \cdot z_2^2
+ \rfrac{1}{2}\cdot (\sigma^2 +a_3\sigma^3 + a_4\sigma^4 ) \cdot [\II_2''(-z_0)+ \\
&&+ \II_2'''(-z_0)\cdot (-z_2)]
+ \rfrac{1}{6}\cdot (\sigma^3 \mu_3 + b_4 \sigma^4) \cdot \II_3'''(-z_0)
+ \rfrac{1}{24}\cdot \sigma^4 \mu_4 \cdot\II_4''''(-z_0) +o(\sigma^4)\, ,
\end{eqnarray*}
where $a_3 = \mu_3$, $a_4=\left(\mbox{$\frac{7}{12}$}\mu_4 - \mbox{$\frac{5}{4}$}\right)$  and $b_4=\mbox{$\frac{3}{2}$}(\mu_4-1) $, i.e.~equal to the third and fourth order terms of the expansion \eqref{eq:MomentsXlognormExpans}.
(Note that if $\sigma = \sigma_{N}$ then $a_3=a_4=b_4 = 0$.)
We observe $\II_j' = -(\phi-id)^j f_L$ and
\begin{equation}
\II_j'' = j(\phi-id)^{j-1} f_L - (\phi-id)^j f_L'
= -  j\II_{j-1}' - (\phi-id)^j f_L' \, .
\label{eq:help_jKjmin1Prim}
\end{equation}
Setting the second order terms in the above equation equal to zero we recover $z_2 = -\frac{\sigma^2}{2 f_L(\qL)}\cdot  \II_2''(q)
= \frac{\sigma^2}{2 f_L(\qL)}\cdot  \big((\phi-id)^2 f_L\big)'$, which is the one-dimensional variant of Theorem \ref{thm:PerturbVarMultiDim}.
Setting the third order terms equal to zero leads
$z_3 = -\frac{\sigma^3}{6 f_L(\qL)}\cdot (3 \cdot a_3 \cdot \II_2''(q) + \mu_3\cdot\II_3'''(q))
= \frac{\sigma^3\mu_3}{6 f_L(\qL)}\cdot  [(\phi-id)^3 f_L']'(q)$,
where the second equation follows from \eqref{eq:help_jKjmin1Prim}.
Setting the fourth order term equal to zero we obtain
%
$$0=f_{L}(q) z_4 +\sigma^4 \left[
- \frac{f_{L}'\II_2''^2}{8 f_L^2}
+ \frac{a_4 \II_2''}{2}
+ \frac{\II_2''' \II_2''}{4f_L}
+ \frac{b_4  \II_3'''}{6}
+ \frac{\mu_4 \II_4''''}{24}\right]\!(q) \, .$$
Observing that $\left(\frac{{\II_2''}^2}{f_L}\right)' = -\frac{f_L'{\II_2''}^2}{f_L^2} + 2\frac{{\II_2''} \II_2'''}{f_L}$ we derive
\begin{eqnarray*}
z_4 &=&  -\frac{\sigma^4}{24 f_{L}(q)}\cdot \left[ \mu_4 \II_4''' +  3\frac{\II_2''^2}{f_L}
+ 12{a_4 \II_2'} + 4{b_4  \II_3''} \right]'(q)\\
&=& -\frac{\sigma^4}{24 f_{L}(q)}\cdot \left[  -\mu_4[(\phi-id)^4 f_L']'
+  3\frac{\II_2''^2}{f_L}
+ (7\mu_4-15) \II_2' + (6\mu_4 -4\mu_4-6)  \II_3'' \right]'(q)\\
&=& -\frac{\sigma^4}{24 f_{L}(q)}\cdot \bigg[- \mu_4[(\phi-id)^4 f_L']'
+ 3\frac{\II_2''^2}{f_L}
+ (7\mu_4-15-3(2\mu_4-6)) \II_2'\\
&&- (2\mu_4 -6) (\phi-id)^3 f_L' \bigg]'(q)\\
&=& \frac{\sigma^4}{24 f_{L}(q)}\cdot \left[ \mu_4[(\phi-id)^4 f_L']'
-  3\frac{\II_2''^2}{f_L}
- (\mu_4 +3) \II_2' + (2\mu_4 -6) (\phi-id)^3 f_L' \right]'(q) \, ,
\end{eqnarray*}
where the second and third equality follow again from \eqref{eq:help_jKjmin1Prim}, which proofs the fourth order expansion; hence part a) is proved.

Ad b): Let's turn to the expression for $\phi^*$:
setting $\psi = \phi-\qL$, we can rewrite
the value-at-risk in third order expansion of part a) when  performing the differentiation
\begin{eqnarray*}
\VaR_\alpha[S(\phi)] &=& \qL
-  \frac{1}{f_L(q)} \cdot \Big\{
 \big(\psi^2 f_L'(q)-2\psi f_L(q)  \big) \cdot \frac{\sigma_{lN}^2}{2}
+ \big(\psi^3 f_L''(q) -3\psi^2 f_L'(q) \big) \cdot \frac{\sigma_{lN}^3 \mu_3}{6} \Big\} +   o(\sigma_{lN}^3)\\
&=& (a/3) 
\cdot \psi^3 + (b/2)
\cdot \psi^2 + c\cdot \psi + q +   o(\sigma_{lN}^3)\, ,
\end{eqnarray*}
with $a = -(\mu_3\sigma_{lN}^3/2) \cdot (f_L''/f_L)(q)$, 
$b= (\mu_3\sigma_{lN}-1)\sigma_{lN}^2\cdot (f_L'/f_L)(q)$, and $c = \sigma_{lN}^2$. 
Setting the differential with respect to $\psi$ equal to zero yields the quadratic formula 
which is solved by $\psi_{\pm} = (-b \pm \sqrt{b^2-4ac})/(2a)$.
Only $\psi_+$ constitutes a (local) minimum of the third order polynomial in $\psi$,
since its second order derivative evaluated at $\psi_\pm$ reads
$2a\psi_\pm +b = \pm \sqrt{b^2-4ac}$ which is only positive for $\psi_+$.
Hence the locally minimal $\phi$ is given by $\phi^* = \qL+\psi_+$.
Inserting the parameters $a,b$, and $c$ and straight forward calculus leads the assertion.

\end{appendix}

\end{document}